\documentclass[%
  reprint,
  amsmath,
  amssymb,
  aps,
  floatfix,
]{revtex4-2}

\usepackage{graphicx}
\usepackage{dcolumn}
\usepackage{bm}
\usepackage{hyperref}
\usepackage{xcolor}
\usepackage{subcaption}

\usepackage{booktabs}
\usepackage{newtxtext,newtxmath}

\begin{document}

\title{Constrained Optimization of Higher-Order Cluster-Expansion Hamiltonians for Alloys Using Simulated Bifurcation}

\author{Kazuhide Ichikawa}
\email{ichikawa.kazuhide@jp.panasonic.com}
\affiliation{Technology Sector, Panasonic Holdings Corporation, 1006 Kadoma, Kadoma City, Osaka 571-8508, Japan}
\affiliation{Graduate School of Engineering, The University of Osaka, 2-1 Yamadaoka, Suita, Osaka 565-0871, Japan}

\author{Satoru Ohuchi}
\email{ohuchi.s@jp.panasonic.com}
\affiliation{Technology Sector, Panasonic Holdings Corporation, 1006 Kadoma, Kadoma City, Osaka 571-8508, Japan}

\author{Tomoyasu Yokoyama}
\email{yokoyama.tomoyasu@jp.panasonic.com}
\affiliation{Technology Sector, Panasonic Holdings Corporation, 1006 Kadoma, Kadoma City, Osaka 571-8508, Japan}

\author{Takuma Saito}
\email{takuma.saito@fixstars.com} 
\affiliation{Fixstars Amplify Corporation, 1-1-1 Shibaura, Minato-ku, Tokyo 105-0023, Japan}

\author{Yoshiki Matsuda}
\email{y\_matsuda@fixstars.com} 
\affiliation{Fixstars Amplify Corporation, 1-1-1 Shibaura, Minato-ku, Tokyo 105-0023, Japan}
\affiliation{Fixstars Corporation, 1-1-1 Shibaura, Minato-ku, Tokyo 105-0023, Japan}


\begin{abstract}
  Identifying ground-state and low-energy atomic configurations is a central problem in alloy design.
  The cluster-expansion (CE) method represents configurational energetics on a fixed lattice as an effective Hamiltonian; for binary alloys, higher-order CE models become polynomial Ising Hamiltonians.
Using Au--Cu as a model binary alloy, we formulate cubic and quartic
cluster-expansion Hamiltonians as penalty-augmented polynomial
unconstrained binary optimization (PUBO) problems under fixed-composition
constraints. We optimize these PUBO problems using SQBM+, a
simulated-bifurcation-based solver that can treat higher-order polynomial
binary objectives directly. 
This direct PUBO treatment avoids the need to construct an explicit
quadratic reformulation with auxiliary variables.
Composition constraints are imposed through quadratic penalty terms,
whose weights are estimated from derivative coefficients of the
continuous relaxation of the CE objective.
  Benchmark calculations for systems up to 2048 atoms show that SQBM+ robustly obtains low-energy feasible configurations for cubic CE models and remains effective for many quartic instances.
  Formation-energy convex hulls constructed from the optimized configurations recover the CuAu and Cu$_3$Au ordering trends and reveal finite-size effects at off-stoichiometric compositions.
  These results demonstrate simulated bifurcation as a practical route to constrained higher-order CE optimization for alloy configuration search.
\end{abstract}
\maketitle

\section{Introduction}
\label{sec:introduction}

Ising machines, both analog and digital, have emerged as specialized optimizers for combinatorial problems and have been applied to a wide range of real-world tasks via quadratic unconstrained binary optimization (QUBO) formulations \cite{Mohseni2022}. In materials science, one particularly active research direction treats structure determination in systems with large configurational degrees of freedom, such as proteins and crystals, as an energy-minimization problem that can be mapped onto either an Ising (quadratic spin) or a QUBO (quadratic binary) formulation.

On the molecular side, early studies formulated lattice protein folding for quantum annealers \cite{PerdomoOrtiz2012}, and subsequent work established folding and design benchmarks on contemporary hardware \cite{Irback2022,Irback2024}. Beyond biomacromolecules, digital annealing has been integrated into computer-aided structure elucidation by formulating scaffold substituent selection as a QUBO subproblem \cite{Lee2024}, while annealing-based formulations have also been used to efficiently search multimolecular adsorption configurations relevant to heterogeneous catalysis \cite{Sampei2023}. In parallel, higher-order unconstrained binary optimization (HUBO) with tensor-train solvers has been proposed for adsorbate configurations on alloy surfaces, underscoring the importance of many-body terms in physically realistic adsorption models \cite{Do2025}.

On the crystalline side, progress has spanned both methods and applications. For crystal structure prediction (CSP), many-body, especially three-body, interactions have been incorporated into Ising/QUBO (and more generally HUBO) encodings \cite{Couzinie2024}, while a machine-learning loop with annealing hardware (FMQA: factorization-machines-with-quantum-annealing) \cite{Kitai2020} has been used to address grand-canonical CSP by fitting a factorization machine and casting its quadratic surrogate directly as a QUBO \cite{CouzinieML2025}. In parallel, Liang \emph{et al.} developed CRYSIM (``CRYstal structure prediction with Symmetry-encoded Ising Machine''), which encodes space-group symmetry and Wyckoff positions to shrink the search space \cite{CRYSIM2025}. Complementing these approaches, an integer-programming framework provides formal \emph{optimality guarantees} and has been validated not only with Gurobi but also on D-Wave hardware \cite{Gusev2023}. Applications now include alloy and oxide-catalyst materials spaces via a quantum-inspired cluster-expansion method \cite{Choubisa2023}, 
battery materials, where quantum annealing has been used to sample ionic ground states using a grand-canonical transformation of the Coulomb-energy cost function \cite{Binninger2024}, 
and high-entropy alloys, where an active-learning workflow dubbed QALO (``quantum annealing--assisted lattice optimization'') reproduces segregation trends \cite{Xu2025}; relatedly, samplers based on quantum annealing enable scalable exploration of low-energy configurations and thermodynamics in disordered materials \cite{Camino2025}.

Here, we focus on site-occupancy configurations in crystals. Our prior work showed that mapping cluster-expansion (CE) energetics to QUBO enables efficient configuration search in an Au--Cu lattice, yielding plausible solutions under composition constraints, in agreement with experimental phase-stability trends \cite{Ichikawa2024}. Leveraging recent hardware and software advances (e.g., third-generation digital annealers \cite{fujitsuDAUserGuide2024} with $\sim 10^5$ fully connected bits), we solved instances with up to 16384 atoms within minutes and reached models of $\sim 6$\,nm, suggesting a route toward bridging atomic- and macroscale phenomena \cite{Ichikawa2024}. 

A remaining challenge in applying Ising-machine-based optimization
to realistic alloy Hamiltonians is the treatment of higher-order
cluster interactions under composition constraints. 
Whereas quadratic CE models are naturally compatible with QUBO
solvers, applying conventional QUBO solvers to cubic and quartic CE
Hamiltonians requires quadratization with auxiliary variables and
additional penalty terms.
Here, we instead provide these Hamiltonians directly to SQBM+ as PUBO models, without
constructing an explicit QUBO reformulation, and benchmark the
optimization of composition-constrained Au--Cu configurations
containing up to 2048 atoms.

Specifically, we employ the polynomial unconstrained binary optimization (PUBO) solver in Toshiba Digital Solutions Corporation's SQBM+ \cite{goto2019combinatorial,goto2021high,kanao2023higherorder}, which provides a GPU-accelerated implementation of the simulated bifurcation (SB) algorithm.
The cubic and quartic CE Hamiltonians are transformed from Ising spin variables to binary variables and optimized as PUBO objectives.
Because SQBM+'s PUBO solver does not natively support constraints, we incorporate penalty functions into the objective function to enforce the target composition.
To obtain low-energy feasible solutions, we estimate appropriate penalty weights from derivative coefficients of the continuously relaxed CE objective.

The remainder of this paper is organized as follows. In Sec.~\ref{sec:ce}, we outline the CE method used for the Au--Cu alloy system and show how the resulting energy function can be cast into a form suitable for optimization with Ising-machine-based solvers. In Sec.~\ref{sec:methods}, we formulate the structural optimization problem using SB, including the treatment of composition constraints by penalty functions and the estimation of appropriate penalty weights. In Sec.~\ref{sec:results}, we present the computational results for Au--Cu alloys over a range of compositions and assess both solution quality and computational performance. Finally, Sec.~\ref{sec:conclusion} summarizes the main findings and discusses future directions.

\section{Cluster Expansion of Au--Cu Alloys}
\label{sec:ce}

In this section, we briefly summarize the procedure used to construct
the CE models adopted in this study.
The overall methodology follows our previous work~\cite{Ichikawa2024},
which employed a quadratic CE model.
Here, we construct cubic and quartic CE Hamiltonians using the same
overall computational workflow.
For detailed theoretical background and practical implementations of the CE approach, readers are referred to Refs.~\cite{SANCHEZ1984334, de1994solid, ceder2000first, wu2016cluster, barroso2022cluster} and to publicly available CE packages such as ATAT, CLUPAN, CELL, CLEASE, ICET, and CASM~\cite{van2009multicomponent, seko2009cluster, troppenz2017predicting, chang2019clease, aangqvist2019icet, puchala2023casm}.

The CE method exploits the fact that atomic sites in a crystal are located on fixed lattice points, thereby enabling an efficient representation of the system energy. For a crystal with $N$ sites and $M$ atomic species, the structure can in general be described by a configuration vector $\vec{\xi}$ consisting of $N$ discrete variables, for example $\xi_i \in \{0,1,\ldots,M-1\}$. 
The CE energy per lattice site, \(E\), is then written as a linear expansion in cluster basis functions $\varphi_\alpha(\vec{\xi})$:
\begin{eqnarray}
E(\vec{\xi}) = \sum_{\alpha} V_{\alpha}\,\varphi_{\alpha}(\vec{\xi}),
\label{eq:E_CE}
\end{eqnarray}
where $\alpha$ labels the clusters and $V_{\alpha}$ denotes the effective cluster interactions (ECIs). 
The ECIs are obtained by fitting to the first-principles energies of
a set of ordered structures, normalized per lattice site in the CE
construction.
For the fully occupied substitutional lattice considered here, the
energy per lattice site is equivalent to the energy per atom.
Once determined, the ECIs can be used to predict the energetics of
atomic configurations in much larger supercells than those included
in the fitting data.

For the binary Au--Cu system considered here, we introduce Ising spin variables $\sigma_i \in \{-1,1\}$, where $\sigma_i=+1$ and $\sigma_i=-1$ denote Au and Cu occupation at site $i$, respectively. In terms of these spin variables, Eq.~\eqref{eq:E_CE} can be rewritten as a polynomial:
\begin{eqnarray}
E(\vec{\sigma}) &=& A + \sum_i B_i \sigma_i + \sum_{i<j} C_{ij} \sigma_i \sigma_j + \sum_{i<j<k} D_{ijk} \sigma_i \sigma_j \sigma_k \nonumber \\
&& + \sum_{i<j<k<l} E_{ijkl} \sigma_i \sigma_j \sigma_k \sigma_l + \cdots,
\label{eq:E_CE_highorder}
\end{eqnarray}
where $A$, $B_i$, $C_{ij}$, $D_{ijk}$, and $E_{ijkl}$ represent the constant, on-site, pair, triplet, and quartet terms, respectively. 
In the present study, we consider CE Hamiltonians containing terms up
to third and fourth order in Eq.~\eqref{eq:E_CE_highorder}, corresponding to maximum cluster
sizes of three and four sites, respectively. These models are hereafter
referred to as the cubic and quartic CE Hamiltonians.

For the determination of ECIs $V_{\alpha}$, we employed the Alloy Theoretic Automated Toolkit (ATAT)~\cite{van2009multicomponent}. First-principles calculations were performed using the Vienna \textit{Ab Initio} Simulation Package (VASP)~\cite{kresse1993ab, kresse1994ab, kresse1996efficiency, kresse1996efficient, kresse1999ultrasoft}. The Projector Augmented Wave (PAW) method with the Perdew--Burke--Ernzerhof (PBE) functional~\cite{Perdew1996} was used. The plane-wave cutoff energy was set to 355\,eV, and a $\Gamma$-centered $k$-point mesh with a spacing of 0.02\,\AA$^{-1}$ was adopted. Atomic positions and lattice constants were relaxed until the residual forces were less than 0.02\,eV\,\AA$^{-1}$. As the parent lattice for the CE, we adopted a face-centered cubic (fcc) lattice with a lattice constant of 3.8\,\AA\ and four atoms per unit cell.

The cubic and quartic CE Hamiltonians were fitted independently using
ATAT; the cubic model was not obtained by truncating the quartic model.
Counting symmetry-inequivalent clusters and excluding the empty and
point clusters, the cubic model contained six pairs and seven triplets,
whereas the quartic model contained six pairs, twenty triplets, and
three four-site clusters.
In both models, the pair clusters extended through the sixth-neighbor
shell (6.582\,\AA).
The maximum triplet diameters were 4.654 and 6.008\,\AA\ for the cubic
and quartic models, respectively, and the maximum diameter of the
four-site clusters in the quartic model was 3.800\,\AA.

The resulting cubic and quartic CE Hamiltonians were used as the
objective functions in the optimization calculations described below.
After transformation from Ising spins to binary variables, they are
represented as cubic and quartic PUBO models, respectively.

\section{Structural Optimization Using Simulated Bifurcation}
\label{sec:methods}

\subsection{Simulated Bifurcation with Constraints}

Here, we present our application of SB to the structural optimization of Au--Cu alloys.
SB is a quantum-inspired algorithm that solves combinatorial optimization problems by simulating the adiabatic evolution of a nonlinear Hamiltonian system.
For a given Ising-type Hamiltonian $H_{\rm Ising}(\bm{\sigma})$, SB considers a parametrically driven $N$-particle Kerr-oscillator system and searches for low-energy spin configurations through its bifurcation dynamics.
The dynamics are governed by a system of differential equations derived from a classical Hamiltonian $H_{\rm SB}(\bm{x}, \bm{y}, t)$, with the continuous dynamics ultimately bifurcating into two states for each degree of freedom, corresponding to the binary spin values \cite{goto2016bifurcation,goto2019combinatorial}.

To execute SB, we utilize the Toshiba SQBM+ platform \cite{goto2021high}.
SQBM+ provides three solvers: QUBO (Quadratic Unconstrained Binary Optimization), QPLIB (Quadratic Programming), and PUBO (Polynomial Unconstrained Binary Optimization).
The QUBO and QPLIB solvers are restricted to quadratic objective functions.
Since the cluster-expansion Hamiltonian for the alloy system contains interactions up to the fourth order, we adopt the PUBO solver, which supports objective functions of up to quartic order.
In the implementation, the physical Ising variables 
\((\sigma_i \in \{-1,1\})\) are represented by binary variables 
\((q_i \in \{0,1\})\) through the transformation 
\(\sigma_i(\bm{q})=2q_i-1\). 
We denote the corresponding binary-variable 
form of the CE Hamiltonian in Eq.~\eqref{eq:E_CE_highorder} by 
\(H(\bm{q})\equiv E(\boldsymbol{\sigma}(\bm{q}))\). 
We employ the Amplify SDK \cite{fixstarsamplifysdkdoc} to manage this transformation and to generate the corresponding penalty function.
As the PUBO solver does not natively support hard constraints, we introduce a penalty function to enforce the compositional constraint on the Au--Cu ratio.
For all compositions considered in this work, $rN$ is an integer.
For an $N$-atom system with a target Au concentration $r$ ($0 \leq r \leq 1$), the penalty term is defined as:
\begin{align}
  P_r(q_0,\dots,q_{N-1}) = \left( rN - \sum_{i=0}^{N-1} q_i \right)^2.
  \label{eq:penalty_formula}
\end{align}
The resulting effective objective function (Hamiltonian) to be minimized is:
\begin{align}
  F_r (\bm{q}) = H(\bm{q}) + w_r P_r (\bm{q}).
  \label{eq:penalized_objective}
\end{align}
Here, \(H(\bm{q})\) is the CE energy per atom expressed in terms of the
binary variables. Since \(P_r(\bm{q})\) is dimensionless, both the
penalty weight \(w_r\) and the penalized objective
\(F_r(\bm{q})\) are expressed in eV/atom.
The choice of $w_r$ is critical: it must be sufficiently large to enforce the constraint ($P_r = 0$), but if set too high, the energy landscape becomes dominated by the penalty term, potentially trapping the solver in local minima with high CE energy $H(\bm{q})$.
Identifying the optimal trade-off for $w_r$ is non-trivial and often solver-dependent.

\subsection{Estimation of Penalty Weight}

One strategy for determining the penalty weight $w_r$ is to scale it
based on the coefficients of the objective function. 
For instance, in a QUBO formulation of the Traveling Salesman
Problem, a sufficient penalty scale can be specified using the maximum
edge weight, ensuring that any global minimizer has a zero constraint penalty \cite{lucas2014ising}.
However, for higher-order polynomial problems,
there is still no generally applicable prescription for choosing an
appropriate penalty weight.

We hypothesize that the local response of the objective function to
variable updates---averaged across the system---provides a practical
scale for identifying an appropriate weight. We estimate the initial weight scale by balancing this response
against the restoring derivative contribution from the composition
penalty at the smallest nonzero composition violation, corresponding
to a one-atom deviation from the target composition.
 Although the variables $q_i$ are binary in the
final optimization problem, the PUBO objective $H(\bm{q})$ is a
polynomial and can be formally regarded as a function of continuous
variables. We therefore define the local objective sensitivity using
the differential coefficient of this continuous extension rather than
by a finite difference associated with flipping a binary variable.
Let
\begin{align}
  g_i^H(\bm{q}) = \left.\frac{\partial H}{\partial q_i}\right|_{\bm{q}}
\end{align}
denote the derivative coefficient of the objective function evaluated at a binary configuration $\bm{q}$.
This quantity corresponds to the coefficient of the infinitesimal change in the relaxed polynomial objective and is used as a practical proxy for the local driving force acting on each variable.

To obtain a system-averaged measure of the objective response rather
than a worst-case coefficient, we employ a two-stage averaging procedure.
First, we compute the spatial average of the derivative coefficients over all $N$ variables for a given configuration $\bm{q}$:
\begin{align}
  \overline{g_H(\bm{q})} = \frac{1}{N} \sum_{i=0}^{N-1} g_i^H(\bm{q}).
\end{align}
Second, since these values depend on the specific configuration $\bm{q}$, we compute the statistical average over an ensemble of $K$ random configurations $\{\bm{q}^{(k)}\}_{k=1}^K$, denoted as $\langle \cdot \rangle$:
\begin{align}
  \langle \overline{g_H} \rangle = \frac{1}{K} \sum_{k=1}^{K} \overline{g_H(\bm{q}^{(k)})}.
  \label{eq:averaged_objective_sensitivity}
\end{align}
In the analysis below, this statistical average is evaluated over random configurations satisfying the target composition constraint.

For the quadratic composition penalty in Eq.~\eqref{eq:penalty_formula}, the corresponding derivative coefficient is
\begin{align}
  g_i^P(\bm{q}) = \left.\frac{\partial P_r}{\partial q_i}\right|_{\bm{q}} = 2\left(\sum_{j=0}^{N-1} q_j - rN\right).
\end{align}
This derivative vanishes for configurations that exactly satisfy the
composition constraint and has magnitude 2 at the smallest nonzero
composition violation, corresponding to a one-atom deviation from the
target composition. Accordingly, the derivative contribution from the
weighted penalty term $w_rP_r$ has magnitude $2w_r$ at this deviation.
We use the comparison between this penalty response and the averaged
objective response as a heuristic guide for estimating the initial
weight scale. This comparison motivates the proportionality
$w_r \propto |\langle\overline{g_H}\rangle|$.
Absorbing the factor of 2 and the additional empirical adjustment into
a correction factor $C_r$, we estimate the initial value of $w_r$ as
\begin{equation}
  w_r \approx
  C_r\left|\langle\overline{g_H}\rangle\right|,
\end{equation}
where $C_r$ is calibrated empirically from the convergence tests
described in Sec.~\ref{sec:results}.

\section{Results}
\label{sec:results}

\subsection{Cluster-Expansion Setup}

  Using the cubic and quartic CE Hamiltonians constructed in Sec.~\ref{sec:ce},
  we generated benchmark optimization instances for periodic Au--Cu supercells
  with $N=32, 256$, and $2048$ atoms.
  As a preliminary check of the CE Hamiltonians and the SQBM+ implementation,
  we first performed unconstrained SQBM+ calculations.
  These calculations consistently found the lowest-energy configuration at $r=0.5$
  for all supercell sizes.
The corresponding ordered structure is consistent with that reported in our previous study~\cite{Ichikawa2024}.

  The composition-constrained instances generated from the same CE Hamiltonians
  are used in the penalty-weight analysis and solver benchmarks described below.

\begin{figure*}[htbp]
  \centering
  \begin{subfigure}[t]{0.48\textwidth}
    \centering
    \includegraphics[width=\linewidth]{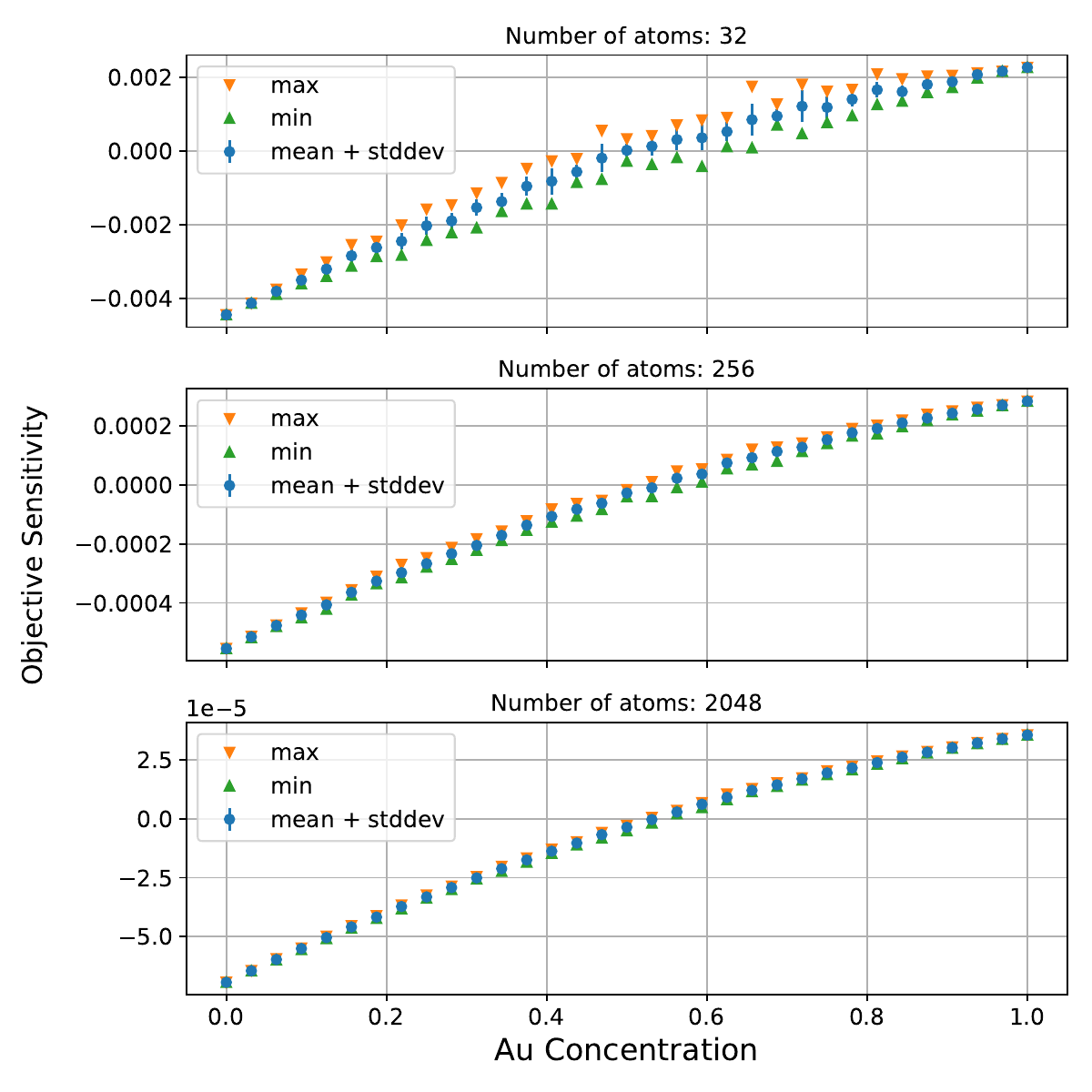}
    \caption{Cubic CE model.}
    \label{fig:differential_cubic}
  \end{subfigure}
  \hfill
  \begin{subfigure}[t]{0.48\textwidth}
    \centering
    \includegraphics[width=\linewidth]{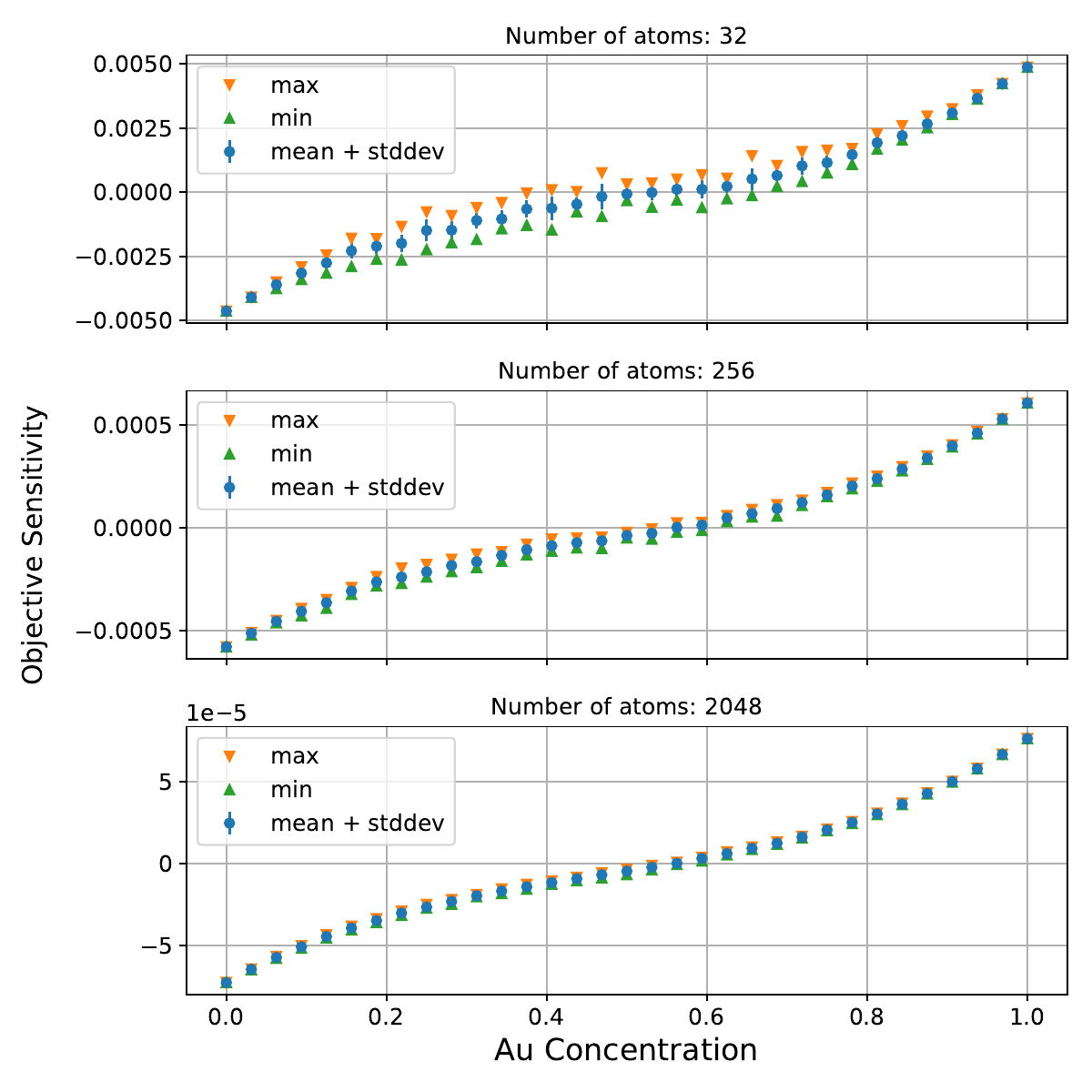}
    \caption{Quartic CE model.}
    \label{fig:differential_quartic}
  \end{subfigure}
  \caption{
Statistics of the objective sensitivity $\overline{g_H(\bm{q})}$ evaluated over 10 random configurations satisfying the target composition constraint for  (a) the cubic CE model and (b) the quartic CE model. 
The mean values correspond to
\(\langle\overline{g_H}\rangle\) in
Eq.~\eqref{eq:averaged_objective_sensitivity}.
Error bars indicate one standard deviation, and the maximum and
minimum values show the full range among the sampled configurations.}
  \label{fig:differential}
\end{figure*}

\subsection{Empirical Calibration of the Penalty-Weight Scale}

To empirically calibrate the correction factor \(C_r\) in the
penalty-weight estimate proposed in Sec.~\ref{sec:methods}, we evaluated
the averaged derivative coefficient of the objective function,
\(\langle \overline{g_H} \rangle\), over 10 randomly sampled
configurations satisfying the composition constraint.

Figures~\ref{fig:differential_cubic} and \ref{fig:differential_quartic} show these averaged derivative coefficients for the cubic and quartic CE Hamiltonians, respectively.
In both cases, the magnitude of the coefficient decreases approximately as \(1/N\).
Notably, the coefficient becomes very small and reaches a minimum around  $r=0.5$, which is consistent with the existence of the energy minimum at this concentration.
A key difference appears in the dependence on $r$:
the derivative coefficient derived from the cubic CE Hamiltonian shows a near-linear relationship,
whereas that derived from the quartic CE Hamiltonian exhibits a pronounced cubic dependence.
This arises from the algebraic structure of the cluster-expansion; in the cubic Hamiltonian, the quadratic terms in the local field effectively average to zero, leaving the linear behavior dominant, whereas the quartic Hamiltonian retains higher-order contributions.

\begin{figure*}[htbp]
  \centering
  \begin{subfigure}[t]{0.48\textwidth}
    \centering
    \includegraphics[width=\linewidth]{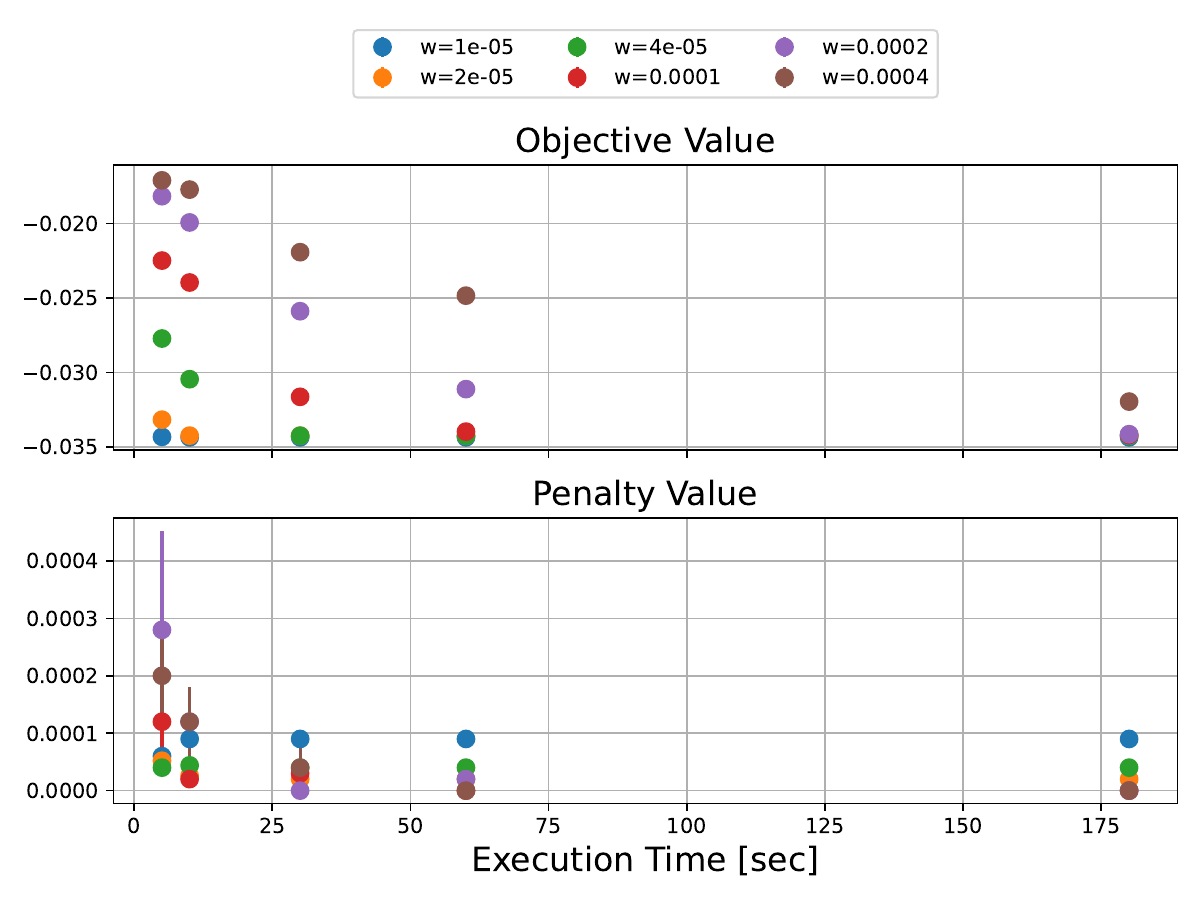}
    \caption{SQBM+.}
    \label{fig:sqbm_convergence_cubic}
  \end{subfigure}
  \hfill
  \begin{subfigure}[t]{0.48\textwidth}
    \centering
    \includegraphics[width=\linewidth]{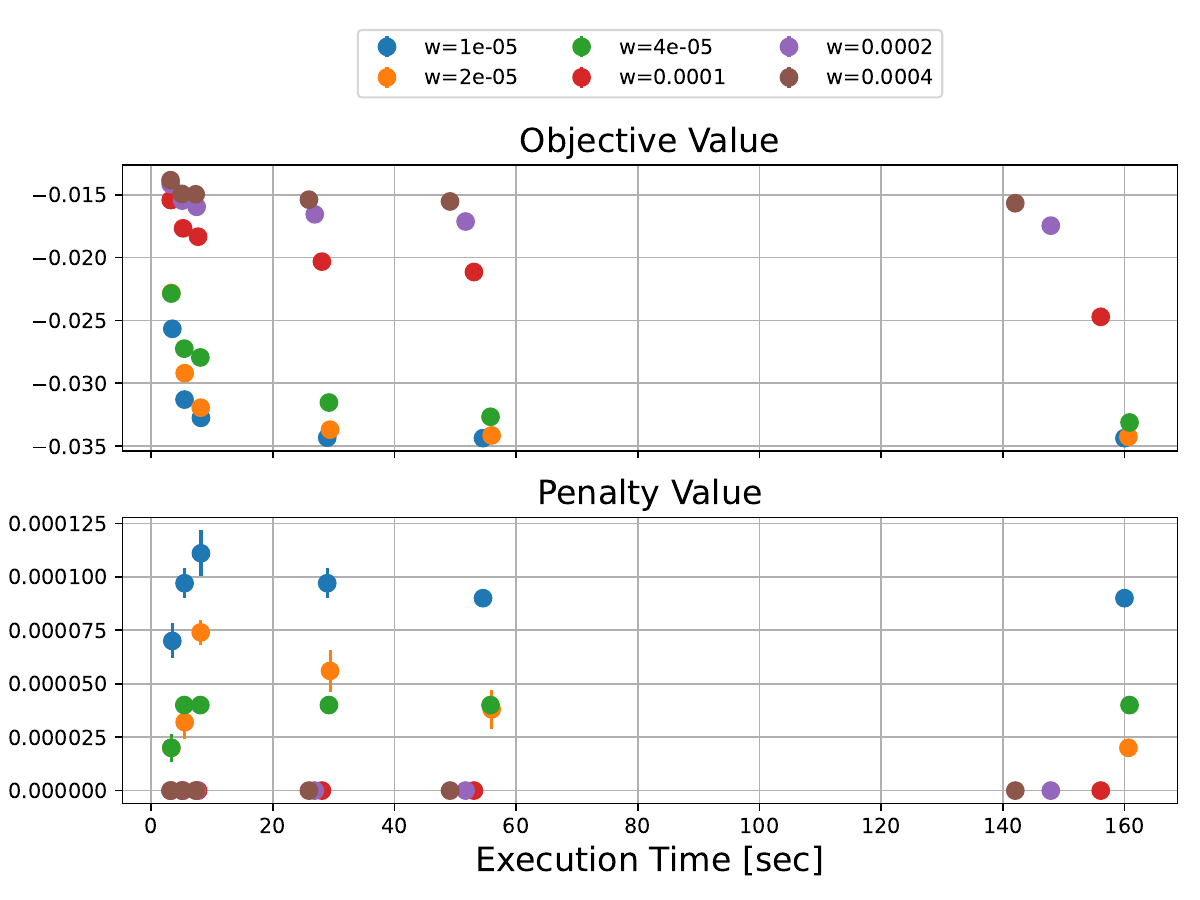}
    \caption{OpenJij.}
    \label{fig:openjij_convergence_cubic}
  \end{subfigure}
  \caption{Convergence of (a) SQBM+ and (b) OpenJij for the cubic CE Hamiltonian at $N=2048$ and $r=0.75$. In each panel, the upper and lower plots show the CE objective $H(\bm{q})$ and the weighted composition-penalty contribution $w_r P_r(\bm{q})$, respectively, as functions of execution time. These two terms constitute the penalized objective $F_r(\bm{q})$ in Eq.~\eqref{eq:penalized_objective}. Colors indicate the tested values of $w_r$. Markers and error bars represent the mean and standard error over 10 runs, respectively.}
  \label{fig:solver_convergence_cubic}
\end{figure*}

Following Sec.~\ref{sec:methods}, we estimated the initial penalty weight as
$w_r \approx C_r|\langle\overline{g_H}\rangle|$.
Because the purpose of this calibration was to determine a practical
order-of-magnitude correction rather than an instance-specific optimum,
we examined the largest system considered, $N=2048$, at the
representative Au-rich composition $r=0.75$, corresponding to the
CuAu$_3$ stoichiometry.
At this composition, the averaged objective derivative remains clearly
nonzero, allowing the balance between objective minimization and
constraint enforcement to be assessed.
Figure~\ref{fig:solver_convergence_cubic} shows the corresponding convergence tests for SQBM+ and OpenJij, an open-source simulated-annealing framework used here as a CPU-based reference solver~\cite{openjij2024}.
For both solvers, weights below $1\times10^{-4}$ frequently led to
constraint violations, whereas weights above $2\times10^{-4}$ slowed
the minimization of the CE objective.
Figure~\ref{fig:differential_cubic} gives
$|\langle\overline{g_H}\rangle|\approx2\times10^{-5}$ at this
composition; thus, the smallest tested weight that reliably suppressed
constraint violations, approximately $1\times10^{-4}$, was about five
times the averaged objective derivative coefficient.
We therefore adopted $C_r=5$ as a common empirical correction factor
for the benchmarks below.

\subsection{Benchmark for Computational Capability}

The benchmark calculations used the PUBO solver of Toshiba SQBM+
v2.0.3, executed on a GPU, and the {\tt SASampler} provided by OpenJij v0.11.6,
executed on a CPU.
For each instance, we performed 100 independent runs with both solvers
using the same penalty weight $w_r$.
SQBM+ was run with wall-clock limits of 10 s for $N=256$ and 180 s for
$N=2048$, whereas OpenJij was run for 20,000 and 5,000 sweeps, respectively.
Table~\ref{tab:benchmark} summarizes the results.

Table~\ref{tab:benchmark} shows that SQBM+ returned the best-found
feasible objective value in at least 99\% of the 100 runs for every
cubic CE instance, whereas the success rates for the quartic
Hamiltonian varied across the benchmark instances, particularly for \(N=2048\).
OpenJij reproduced the SQBM+ reference value for some relatively easy
instances, including $r=0.5$ and several 256-atom cases, but its success
rates were low for most larger or higher-order instances.
Because SQBM+ and OpenJij were run on different hardware and with
different stopping criteria, this benchmark does not separate
differences between the optimization methods from the effects of the
computing hardware.
It nevertheless provides a practical comparison of the two
computational setups used here.
Under these settings, SQBM+ returned the best-found feasible objective
value at least as frequently as OpenJij for every benchmark instance
and more frequently for most of them, with prescribed wall-clock
limits that were comparable to or shorter than the mean runtimes
measured for OpenJij.

\begin{table*}[htbp]
  \centering
\caption{\label{tab:benchmark}
Benchmark results for the cubic and quartic CE Hamiltonians obtained using SQBM+ and OpenJij.
The ``Best Found Objective'' columns give the lowest CE objective among the feasible solutions obtained by each solver.
In the ``Time [s]'' columns, the SQBM+ entries are the prescribed wall-clock limits, whereas the OpenJij entries are the mean wall-clock runtimes per run for the prescribed numbers of sweeps. Neither quantity represents the time required to first reach the best-found objective value.
The ``Rate of Best'' columns give the fraction of all 100 runs that returned a feasible solution reaching the best feasible objective value found by SQBM+ for the same instance.}
  \setlength{\tabcolsep}{5pt}
  \begin{tabular}{cccc c c cc cc}
    \toprule
    & & & & \multicolumn{2}{c}{Best Found Objective} & \multicolumn{2}{c}{Time [s]} & \multicolumn{2}{c}{Rate of Best} \\
    \cmidrule(lr){5-6} \cmidrule(lr){7-8} \cmidrule(lr){9-10}
    Degree & $N$ & $r$ & Weight ($w_r$) & SQBM+ & OpenJij & SQBM+ & OpenJij & SQBM+ & OpenJij \\
    \midrule
    Cubic & 256  & 0.125 & 0.002034 & -0.014264 & -0.009063 & 10 & 12.7 & 1.0 & 0.0  \\
    Cubic & 256  & 0.250 & 0.001334 & -0.050948 & -0.050948 & 10 & 16.3 & 1.0 & 0.75 \\
    Cubic & 256  & 0.375 & 0.000682 & -0.051254 & -0.049018 & 10 & 20.8 & 1.0 & 0.0  \\
    Cubic & 256  & 0.500 & 0.000135 & -0.055128 & -0.055128 & 10 & 22.7 & 1.0 & 1.0  \\
    Cubic & 256  & 0.625 & 0.000374 & -0.049590 & -0.049590 & 10 & 21.8 & 1.0 & 1.0  \\
    Cubic & 256  & 0.750 & 0.000768 & -0.032670 & -0.032670 & 10 & 19.1 & 1.0 & 0.99 \\
    Cubic & 256  & 0.875 & 0.001136 & -0.014655 & -0.011511 & 10 & 14.8 & 1.0 & 0.0  \\
    \midrule
    Cubic & 2048 & 0.125 & 0.000252 & -0.015705 & -0.007054 & 180 & 180.4 & 0.99 & 0.0 \\
    Cubic & 2048 & 0.250 & 0.000166 & -0.050948 & -0.034865 & 180 & 247.2 & 1.0  & 0.0 \\
    Cubic & 2048 & 0.375 & 0.000087 & -0.052146 & -0.044793 & 180 & 297.7 & 1.0  & 0.0 \\
    Cubic & 2048 & 0.500 & 0.000018 & -0.055128 & -0.055128 & 180 & 324.2 & 1.0  & 0.95 \\
    Cubic & 2048 & 0.625 & 0.000046 & -0.049590 & -0.049590 & 180 & 308.7 & 1.0  & 0.95 \\
    Cubic & 2048 & 0.750 & 0.000097 & -0.034202 & -0.033498 & 180 & 264.9 & 1.0  & 0.0 \\
    Cubic & 2048 & 0.875 & 0.000141 & -0.016187 & -0.010093 & 180 & 201.5 & 0.99 & 0.0 \\
    \midrule
    Quartic & 256 & 0.125 & 0.001822 & -0.015047 & -0.013350 & 10 & 37.6 & 0.01 & 0.0 \\
    Quartic & 256 & 0.250 & 0.001071 & -0.047251 & -0.047251 & 10 & 49.2 & 1.0  & 0.07 \\
    Quartic & 256 & 0.375 & 0.000533 & -0.049562 & -0.048342 & 10 & 60.5 & 1.0  & 0.0  \\
    Quartic & 256 & 0.500 & 0.000185 & -0.056137 & -0.056137 & 10 & 64.7 & 1.0  & 1.0  \\
    Quartic & 256 & 0.625 & 0.000246 & -0.049829 & -0.049829 & 10 & 63.8 & 1.0  & 1.0  \\
    Quartic & 256 & 0.750 & 0.000804 & -0.033455 & -0.033455 & 10 & 57.7 & 1.0  & 0.89 \\
    Quartic & 256 & 0.875 & 0.001703 & -0.012581 & -0.009343 & 10 & 45.9 & 1.0  & 0.0  \\
    \midrule
    Quartic & 2048 & 0.125 & 0.000222 & -0.015824 & -0.011797 & 180 & 219.4 & 0.01 & 0.0  \\
    Quartic & 2048 & 0.250 & 0.000132 & -0.047251 & -0.037423 & 180 & 302.0 & 0.75 & 0.0  \\
    Quartic & 2048 & 0.375 & 0.000070 & -0.050628 & -0.044296 & 180 & 368.3 & 1.0  & 0.0  \\
    Quartic & 2048 & 0.500 & 0.000023 & -0.056137 & -0.056137 & 180 & 409.7 & 1.0  & 0.93 \\
    Quartic & 2048 & 0.625 & 0.000031 & -0.049829 & -0.049829 & 180 & 387.2 & 1.0  & 0.86 \\
    Quartic & 2048 & 0.750 & 0.000102 & -0.033633 & -0.026893 & 180 & 326.1 & 0.74 & 0.0  \\
    Quartic & 2048 & 0.875 & 0.000214 & -0.012548 & -0.008483 & 180 & 241.4 & 0.01 & 0.0  \\
    \bottomrule
  \end{tabular}
\end{table*}

\subsection{Formation-Energy Convex Hulls}

To assess the physical plausibility of the low-energy solutions
obtained by composition-constrained SQBM+ optimization, we examined
whether their formation-energy landscape and ordering patterns are
consistent with established knowledge of Au--Cu alloys.
For this purpose, we evaluated 33 uniformly spaced Au concentrations,
\(r=k/32\) (\(k=0,1,\dots,32\)), for system sizes \(N=32\), 256, and
2048, so that each supercell contained \(kN/32\) Au atoms.
For each of the 31 non-endpoint compositions, the initial penalty
weight \(w_r\) was estimated by the derivative-based method described
in Sec.~\ref{sec:methods}, using the correction factor \(C_r=5\).
A preliminary SQBM+ run was then performed with this initial weight.
If it returned an infeasible solution, i.e., one violating the
composition constraint, the penalty weight was doubled and the
preliminary calculation was repeated until a feasible solution was
obtained.
The resulting weight was then held fixed for 10 independent SQBM+
runs.
All 10 runs returned feasible configurations at every non-endpoint
composition.
The wall-clock limits were set to 1~s for \(N=32\), 10~s for
\(N=256\), and 180~s for \(N=2048\); the latter two were the same as
those used for the benchmark calculations in
Table~\ref{tab:benchmark}.
The optimized energy \(E_{\rm opt}(r)\) at each non-endpoint
composition was taken as the lowest CE objective value obtained in
these 10 runs.
The pure-component endpoints \(r=0\) and \(r=1\), which provide the
reference energies \(E_{\rm Cu}\) and \(E_{\rm Au}\), respectively,
were obtained by direct evaluation of the CE Hamiltonian for the
all-Cu and all-Au configurations, without optimization.

The formation energy per atom at a given Au concentration \(r\) was
evaluated from the CE energy of the optimized configuration as
\[
  E_{\rm form}(r)=E_{\rm opt}(r)-rE_{\rm Au}-(1-r)E_{\rm Cu},
\]
where \(E_{\rm opt}(r)\) is the energy per atom of the optimized
configuration at concentration \(r\), and \(E_{\rm Au}\) and \(E_{\rm Cu}\)
are the corresponding energies per atom of the pure Au and Cu
structures. The lower convex envelope of \(E_{\rm form}(r)\) defines the zero-temperature convex hull of the CE Hamiltonian: points on the hull are stable against decomposition within the present CE model, whereas points above it can lower their energy by decomposing into neighboring hull phases.

As a cross-check of the present optimization and energy-evaluation
procedure, we applied SQBM+ to the same quadratic CE Hamiltonian
examined in our previous study, in which optimization was performed
using a Digital Annealer~\cite{Ichikawa2024}.
(The SQBM+ formation-energy profiles are shown in the Supplemental
Material, Fig.~S1.)
The underlying numerical values closely agreed with the previously
reported results over the common compositions and system sizes, with
only very small deviations at some compositions for \(N=2048\).
The same convex-hull vertices were recovered, together with the
associated \(L1_2\)-type Cu$_3$Au and CuAu$_3$ orderings and
\(L1_0\)-type CuAu ordering.

\begin{figure}[htbp]
  \centering
  \begin{minipage}{0.45\textwidth}
    \includegraphics[width=\textwidth]{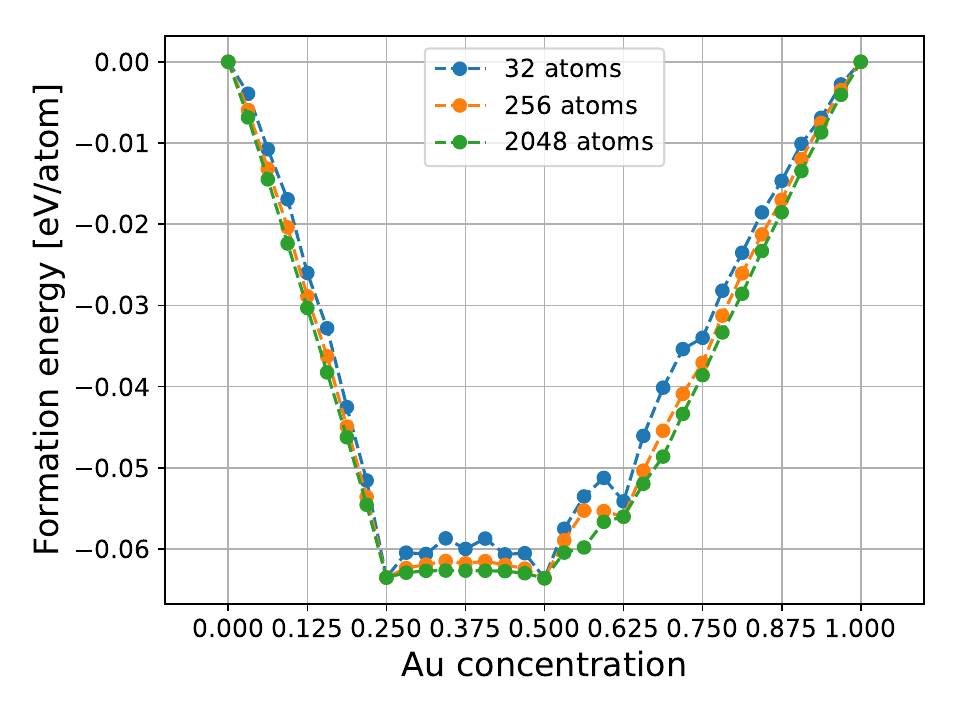}
    \subcaption{Cubic CE model.}
    \label{fig:formation_energy_cubic}
  \end{minipage}
  \hfill
  \begin{minipage}{0.45\textwidth}
    \includegraphics[width=\textwidth]{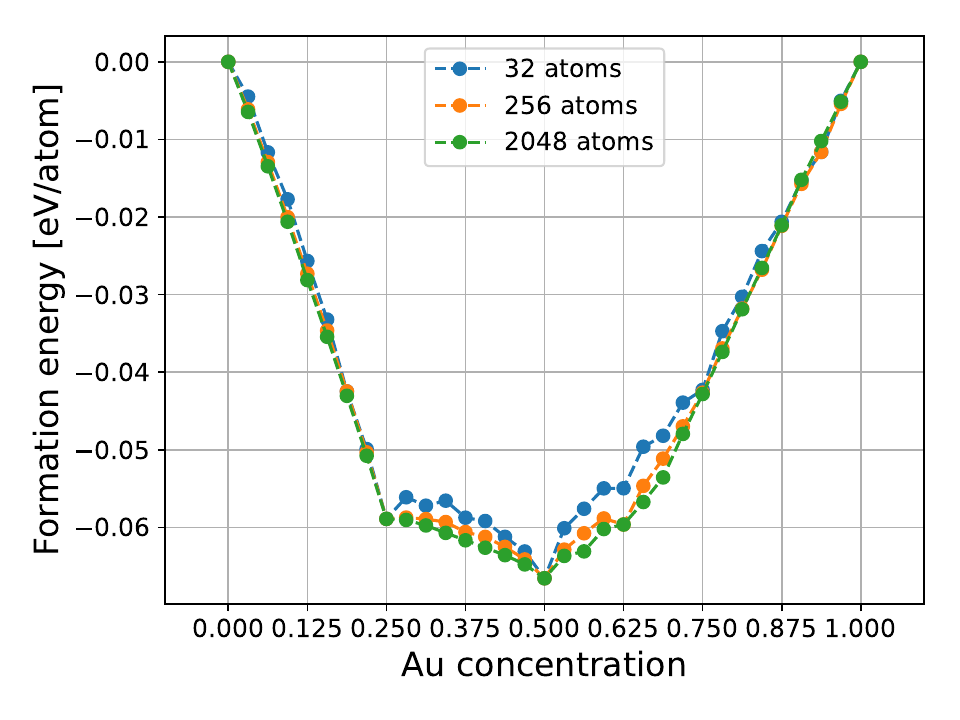}
    \subcaption{Quartic CE model.}
    \label{fig:formation_energy_quartic}
  \end{minipage}
\caption{
  Formation energy per atom as a function of Au concentration \(r\)
  for the optimized configurations obtained with SQBM+:
  (a) cubic and (b) quartic CE Hamiltonians.
  Results are shown for 32-, 256-, and 2048-atom supercells.
  The numerical SQBM+ results used to construct panels (a) and (b)
  are provided in the Supplemental Material,
  Tables~S4--S6 and S7--S9, respectively.
}  \label{fig:formation_energy}
\end{figure}

Figures~\ref{fig:formation_energy_cubic} and
\ref{fig:formation_energy_quartic} show the formation energies of the
optimized configurations obtained from the cubic and quartic CE
Hamiltonians, respectively.
Although the cubic model provides a useful test of direct third-order
PUBO optimization, the comparison with previous Au--Cu studies below
focuses on the quartic CE model, whose inclusion of four-site
interactions permits a more direct comparison with established CE
descriptions of this alloy system~\cite{OzolinsWolvertonZunger1998CuAu,sanati2003adaptive,chang2019clease}.

For the quartic CE Hamiltonian, clear convex-hull vertices appear at
\(r=0.25\) and \(r=0.50\), where the optimized configurations exhibit
\(L1_2\)-type Cu\(_3\)Au and \(L1_0\)-type CuAu ordering, respectively
(see Supplemental Material, Figs.~S2 and S3).
In the present calculations, the latter is also the global
formation-energy minimum.
The occurrence of these two ordered phases is consistent with
experimental phase-diagram and thermodynamic assessments
\cite{OkamotoChakrabartiLaughlinMassalski1987AuCu,
SundmanFriesOates1998AuCu,
FedorovVolkov2016AuCu},
as well as previous first-principles and CE studies
\cite{ZhangKresseWolverton2014CuAu,
OzolinsWolvertonZunger1998CuAu,
sanati2003adaptive,
PANDEY2019151615,
ZhaoXiaZengWang2024CuAu,
LevamakiTianKokkoVitos2018CuAu}.

On the Au-rich side (\(r>0.5\)), the formation-energy landscape is
asymmetric with respect to the Cu-rich side and exhibits a
comparatively shallow convex-hull profile with several weakly
pronounced vertices.
In particular, the \(r=0.75\) point lies on the convex hull for
\(N=32\) and \(256\), whereas it lies slightly above the hull for
\(N=2048\).
Thus, the vertex at \(r=0.75\) is less robust with respect to system
size than those at \(r=0.25\) and \(r=0.50\).
Although the detailed locations of the Au-rich vertices depend on the
electronic-structure approximation and the CE model, the overall
asymmetry and the presence of closely competing low-energy orderings
are qualitatively consistent with trends reported in previous DFT
calculations and DFT-based CE studies
\cite{ZhangKresseWolverton2014CuAu,
OzolinsWolvertonZunger1998CuAu,
sanati2003adaptive,
PANDEY2019151615,
ZhaoXiaZengWang2024CuAu,
LevamakiTianKokkoVitos2018CuAu}.

As a representative example of the Au-rich solutions, we next
examine the optimized configurations at \(r=0.75\)
(see Supplemental Material, Fig.~S4).
For \(N=32\), the optimized configuration consists of a four-plane
repeat comprising one Cu plane and three Au planes and belongs to
space group \(P4/mmm\) (No.~123).
Because the present CE Hamiltonian was derived from GGA calculations
without a Hubbard \(U\) correction, a particularly relevant comparison
is with the uncorrected-GGA results in the CE study of Zhao
et al., which examined both GGA and GGA+\(U\) descriptions.
Their GGA-based CE likewise predicted a \(P4/mmm\)-type CuAu\(_3\)
structure as the zero-temperature ground state at this composition,
whereas the CE based on GGA+\(U\) energies stabilized the
experimentally reported \(L1_2\)-type CuAu\(_3\) phase
\cite{ZhaoXiaZengWang2024CuAu}.
For \(N=256\) and \(2048\), the optimized configurations belong to
space group \(P4/nmm\) (No.~129) and have stacking periods of eight
and sixteen atomic planes, respectively.
Although they belong to the same space group, these two configurations
are distinct long-period structures with different Cu-plane
separations and stacking periods.

The occurrence of such layered and long-period configurations is also
consistent with previous LDA/GGA-based DFT and CE studies of Au-rich
Cu--Au alloys
\cite{OzolinsWolvertonZunger1998CuAu,
sanati2003adaptive,
PANDEY2019151615}.
The progression from four- to eight- and sixteen-plane stacking
periods with increasing system size illustrates that large-cell
optimization can explore long-period configurations that cannot be
represented in smaller cells.
Note that these results do not constitute a formal proof of global optimality
for the largest systems.
The long-period configurations should therefore be interpreted as
low-energy solutions of the present fixed-lattice, GGA-derived CE
Hamiltonian rather than as predictions of experimentally stable phases.

\section{Conclusion}
\label{sec:conclusion}

In this study, we used the PUBO solver of Toshiba SQBM+ for
composition-constrained optimization of cubic and quartic
cluster-expansion Hamiltonians for Au--Cu alloys.
Extending our previous quadratic-QUBO formulation, we formulated the higher-order CE optimization problems directly as PUBO models and solved them without quadratization or auxiliary variables.
Benchmark calculations for systems with up to 2048 atoms demonstrate
that SQBM+ can obtain low-energy feasible configurations for these
direct higher-order representations at system sizes large enough to
examine system-size effects and the emergence of long-period
configurations in the resulting formation-energy convex hulls.

We also developed a practical procedure for estimating the initial
penalty weight by comparing the averaged local response of the CE
objective with the restoring penalty response at the smallest nonzero
composition violation. A common empirical correction factor of
\(C_r=5\) was calibrated from representative convergence tests. 
Combined with iterative weight doubling when necessary, this estimate
yielded feasible solutions throughout the formation-energy
calculations.
This derivative-based estimate therefore provides a
useful initial scale for constrained higher-order CE optimization,
although further testing is needed to establish its transferability to
other Hamiltonians and solvers.

The benchmark calculations showed that SQBM+ repeatedly returned the
best-found feasible objective values for the cubic CE Hamiltonian,
whereas the success rates for the quartic Hamiltonian varied across
the benchmark instances, particularly for \(N=2048\).
Across all benchmark instances, SQBM+ returned the best-found feasible
objective value at least as frequently as OpenJij and more frequently
in most cases, with prescribed wall-clock limits comparable to or
shorter than the mean runtimes measured for OpenJij.
Because the two methods were run on different hardware and with
different stopping criteria, these results should be interpreted as a
practical comparison of the computational setups tested here rather
than as a hardware-normalized comparison of the optimization methods.
The results do not prove global optimality, but show that the SQBM+
PUBO solver provides a practical route to low-energy configurations of
constrained cubic and quartic CE Hamiltonians beyond the range
accessible by direct enumeration.

As a cross-check of the optimization and energy-evaluation
procedure, the quadratic SQBM+ calculations closely reproduced the
formation-energy profiles and principal ordered structures reported 
in our previous study using a Digital Annealer for the same quadratic CE Hamiltonian.
For the quartic CE Hamiltonian,
clear convex-hull vertices and the expected \(L1_2\)-type Cu$_3$Au and
\(L1_0\)-type CuAu ordering were recovered at \(r=0.25\) and \(0.50\),
respectively. The shallow, asymmetric Au-rich landscape exhibited
greater system-size sensitivity: the \(r=0.75\) point lay on the hull
for \(N=32\) and 256 but slightly above it for \(N=2048\), while the
larger cells yielded long-period layered configurations that cannot be
represented in the smallest cell. 
These findings support the physical plausibility of the low-energy solutions obtained for the CE Hamiltonians considered here. 
However, the fidelity of the resulting convex hull depends on the accuracy of the CE models and their underlying first-principles training data.

Future work will extend this framework to multicomponent alloys involving three or more metallic species. Such systems require additional encoding constraints, for example one-hot occupancy constraints at each site together with global composition constraints. Addressing these coupled constraints will require further refinement of the penalty-weight strategy and more systematic benchmarks across different alloy systems, CE models, and solvers.




\bibliography{references_checked_protected_v3.bib}
\end{document}


\title{Supplemental Material for ``Constrained Optimization of
Higher-Order Cluster-Expansion Hamiltonians for Alloys Using
Simulated Bifurcation''}

\author{Kazuhide Ichikawa}
\affiliation{Technology Sector, Panasonic Holdings Corporation,
1006 Kadoma, Kadoma City, Osaka 571-8508, Japan}
\affiliation{Graduate School of Engineering, The University of Osaka,
2-1 Yamadaoka, Suita, Osaka 565-0871, Japan}

\author{Satoru Ohuchi}
\affiliation{Technology Sector, Panasonic Holdings Corporation,
1006 Kadoma, Kadoma City, Osaka 571-8508, Japan}

\author{Tomoyasu Yokoyama}
\affiliation{Technology Sector, Panasonic Holdings Corporation,
1006 Kadoma, Kadoma City, Osaka 571-8508, Japan}

\author{Takuma Saito}
\affiliation{Fixstars Amplify Corporation,
1-1-1 Shibaura, Minato-ku, Tokyo 105-0023, Japan}

\author{Yoshiki Matsuda}
\affiliation{Fixstars Amplify Corporation,
1-1-1 Shibaura, Minato-ku, Tokyo 105-0023, Japan}
\affiliation{Fixstars Corporation,
1-1-1 Shibaura, Minato-ku, Tokyo 105-0023, Japan}

\maketitle

\setcounter{table}{0}
\setcounter{figure}{0}
\renewcommand{\thetable}{S\arabic{table}}
\renewcommand{\thefigure}{S\arabic{figure}}

\section{Complete SQBM+ Results}
\label{sec:supplemental}

Tables~\ref{tab:supp-benchmark-quadratic-32}--\ref{tab:supp-benchmark-quartic-2048}
provide the complete SQBM+ results for the quadratic, cubic, and quartic
cluster-expansion (CE) Hamiltonians at all sampled compositions and
system sizes ($N=32$, 256, and 2048). For each non-endpoint composition,
the listed weight is the final penalty weight held fixed for 10
production runs. The ``Best Found Objective'' is the lowest CE objective
among the feasible solutions obtained in those runs, and the ``Rate of
Best'' is the fraction of the 10 runs that attained that value. The
``Time [s]'' column gives the prescribed SQBM+ wall-clock limit rather
than the time at which the best solution was first found. For the
pure-component endpoints ($r=0$ and 1), the CE energies were evaluated
directly without optimization; therefore, the penalty weight, wall-clock
limit, and rate of best are not applicable and are indicated by dashes.

\begin{table}[p]
  \centering
  \caption{\label{tab:supp-benchmark-quadratic-32} Complete formation-energy results for the quadratic CE Hamiltonian with $N=32$. Column definitions are given in the text.}
  \setlength{\tabcolsep}{16pt}
  \begin{tabular}{cccc ccc}
    \toprule
    Degree & $N$ & $r$ & Weight ($w_r$) & Best Found Objective & Time [s] & Rate of Best \\
    \midrule
    Quadratic & 32 & 0 & --- & 0.005442 & --- & --- \\
    Quadratic & 32 & 0.03125 & 0.023076 & 0.001519 & 1 & 1.0 \\
    Quadratic & 32 & 0.0625 & 0.021812 & -0.003951 & 1 & 1.0 \\
    Quadratic & 32 & 0.09375 & 0.020547 & -0.008187 & 1 & 1.0 \\
    Quadratic & 32 & 0.125 & 0.019282 & -0.013972 & 1 & 1.0 \\
    Quadratic & 32 & 0.15625 & 0.018017 & -0.018522 & 1 & 1.0 \\
    Quadratic & 32 & 0.1875 & 0.016752 & -0.02462 & 1 & 1.0 \\
    Quadratic & 32 & 0.21875 & 0.015488 & -0.029483 & 1 & 1.0 \\
    Quadratic & 32 & 0.25 & 0.014223 & -0.035895 & 1 & 1.0 \\
    Quadratic & 32 & 0.28125 & 0.012958 & -0.03629 & 1 & 1.0 \\
    Quadratic & 32 & 0.3125 & 0.011693 & -0.038232 & 1 & 1.0 \\
    Quadratic & 32 & 0.34375 & 0.010428 & -0.03894 & 1 & 1.0 \\
    Quadratic & 32 & 0.375 & 0.009164 & -0.041197 & 1 & 1.0 \\
    Quadratic & 32 & 0.40625 & 0.007899 & -0.042219 & 1 & 1.0 \\
    Quadratic & 32 & 0.4375 & 0.006634 & -0.044789 & 1 & 1.0 \\
    Quadratic & 32 & 0.46875 & 0.005369 & -0.046124 & 1 & 1.0 \\
    Quadratic & 32 & 0.5 & 0.004104 & -0.049008 & 1 & 1.0 \\
    Quadratic & 32 & 0.53125 & 0.005679 & -0.045875 & 1 & 1.0 \\
    Quadratic & 32 & 0.5625 & 0.00315 & -0.044289 & 1 & 1.0 \\
    Quadratic & 32 & 0.59375 & 0.004959 & -0.041469 & 1 & 1.0 \\
    Quadratic & 32 & 0.625 & 0.00191 & -0.040198 & 1 & 1.0 \\
    Quadratic & 32 & 0.65625 & 0.004439 & -0.037692 & 1 & 1.0 \\
    Quadratic & 32 & 0.6875 & 0.003484 & -0.036734 & 1 & 1.0 \\
    Quadratic & 32 & 0.71875 & 0.004749 & -0.034541 & 1 & 1.0 \\
    Quadratic & 32 & 0.75 & 0.006014 & -0.033897 & 1 & 1.0 \\
    Quadratic & 32 & 0.78125 & 0.007279 & -0.027236 & 1 & 1.0 \\
    Quadratic & 32 & 0.8125 & 0.008544 & -0.022122 & 1 & 1.0 \\
    Quadratic & 32 & 0.84375 & 0.009808 & -0.015774 & 1 & 1.0 \\
    Quadratic & 32 & 0.875 & 0.011073 & -0.010975 & 1 & 1.0 \\
    Quadratic & 32 & 0.90625 & 0.012338 & -0.004941 & 1 & 1.0 \\
    Quadratic & 32 & 0.9375 & 0.013603 & -0.000455 & 1 & 1.0 \\
    Quadratic & 32 & 0.96875 & 0.014868 & 0.005266 & 1 & 1.0 \\
    Quadratic & 32 & 1 & --- & 0.009438 & --- & --- \\
    \bottomrule
  \end{tabular}
\end{table}

\begin{table}[p]
  \centering
  \caption{\label{tab:supp-benchmark-quadratic-256} Complete formation-energy results for the quadratic CE Hamiltonian with $N=256$. Column definitions are given in the text.}
  \setlength{\tabcolsep}{16pt}
  \begin{tabular}{cccc ccc}
    \toprule
    Degree & $N$ & $r$ & Weight ($w_r$) & Best Found Objective & Time [s] & Rate of Best \\
    \midrule
    Quadratic & 256 & 0 & --- & 0.005442 & --- & --- \\
    Quadratic & 256 & 0.03125 & 0.001739 & 0.0011 & 10 & 1.0 \\
    Quadratic & 256 & 0.0625 & 0.001618 & -0.003951 & 10 & 1.0 \\
    Quadratic & 256 & 0.09375 & 0.001497 & -0.008789 & 10 & 1.0 \\
    Quadratic & 256 & 0.125 & 0.001376 & -0.013971 & 10 & 1.0 \\
    Quadratic & 256 & 0.15625 & 0.001255 & -0.018941 & 10 & 1.0 \\
    Quadratic & 256 & 0.1875 & 0.001134 & -0.024619 & 10 & 1.0 \\
    Quadratic & 256 & 0.21875 & 0.001012 & -0.029903 & 10 & 1.0 \\
    Quadratic & 256 & 0.25 & 0.000891 & -0.035895 & 10 & 1.0 \\
    Quadratic & 256 & 0.28125 & 0.00077 & -0.036709 & 10 & 1.0 \\
    Quadratic & 256 & 0.3125 & 0.000649 & -0.038232 & 10 & 1.0 \\
    Quadratic & 256 & 0.34375 & 0.000528 & -0.039903 & 10 & 1.0 \\
    Quadratic & 256 & 0.375 & 0.000407 & -0.042104 & 10 & 1.0 \\
    Quadratic & 256 & 0.40625 & 0.000285 & -0.043181 & 10 & 1.0 \\
    Quadratic & 256 & 0.4375 & 0.000164 & -0.044788 & 10 & 1.0 \\
    Quadratic & 256 & 0.46875 & 0.000172 & -0.046544 & 10 & 1.0 \\
    Quadratic & 256 & 0.5 & 0.000078 & -0.049008 & 10 & 1.0 \\
    Quadratic & 256 & 0.53125 & 0.000199 & -0.046294 & 10 & 1.0 \\
    Quadratic & 256 & 0.5625 & 0.00032 & -0.044289 & 10 & 1.0 \\
    Quadratic & 256 & 0.59375 & 0.000442 & -0.042432 & 10 & 1.0 \\
    Quadratic & 256 & 0.625 & 0.000563 & -0.041105 & 10 & 1.0 \\
    Quadratic & 256 & 0.65625 & 0.000684 & -0.038654 & 10 & 1.0 \\
    Quadratic & 256 & 0.6875 & 0.000805 & -0.036734 & 10 & 1.0 \\
    Quadratic & 256 & 0.71875 & 0.000926 & -0.034961 & 10 & 1.0 \\
    Quadratic & 256 & 0.75 & 0.001047 & -0.033897 & 10 & 1.0 \\
    Quadratic & 256 & 0.78125 & 0.001168 & -0.027655 & 10 & 1.0 \\
    Quadratic & 256 & 0.8125 & 0.00129 & -0.022122 & 10 & 1.0 \\
    Quadratic & 256 & 0.84375 & 0.001411 & -0.016194 & 10 & 1.0 \\
    Quadratic & 256 & 0.875 & 0.001532 & -0.010974 & 10 & 1.0 \\
    Quadratic & 256 & 0.90625 & 0.001653 & -0.005543 & 10 & 1.0 \\
    Quadratic & 256 & 0.9375 & 0.001774 & -0.000454 & 10 & 1.0 \\
    Quadratic & 256 & 0.96875 & 0.001895 & 0.004846 & 10 & 1.0 \\
    Quadratic & 256 & 1 & --- & 0.009438 & --- & --- \\
    \bottomrule
  \end{tabular}
\end{table}

\begin{table}[p]
  \centering
  \caption{\label{tab:supp-benchmark-quadratic-2048} Complete formation-energy results for the quadratic CE Hamiltonian with $N=2048$. Column definitions are given in the text.}
  \setlength{\tabcolsep}{16pt}
  \begin{tabular}{cccc ccc}
    \toprule
    Degree & $N$ & $r$ & Weight ($w_r$) & Best Found Objective & Time [s] & Rate of Best \\
    \midrule
    Quadratic & 2048 & 0 & --- & 0.005442 & --- & --- \\
    Quadratic & 2048 & 0.03125 & 0.000217 & 0.001015 & 180 & 0.10 \\
    Quadratic & 2048 & 0.0625 & 0.000202 & -0.003984 & 180 & 0.10 \\
    Quadratic & 2048 & 0.09375 & 0.000187 & -0.009076 & 180 & 0.90 \\
    Quadratic & 2048 & 0.125 & 0.000172 & -0.014243 & 180 & 1.0 \\
    Quadratic & 2048 & 0.15625 & 0.000157 & -0.01941 & 180 & 1.0 \\
    Quadratic & 2048 & 0.1875 & 0.000142 & -0.02463 & 180 & 0.90 \\
    Quadratic & 2048 & 0.21875 & 0.000127 & -0.030149 & 180 & 1.0 \\
    Quadratic & 2048 & 0.25 & 0.000111 & -0.035895 & 180 & 1.0 \\
    Quadratic & 2048 & 0.28125 & 0.000096 & -0.036956 & 180 & 0.70 \\
    Quadratic & 2048 & 0.3125 & 0.000081 & -0.038684 & 180 & 1.0 \\
    Quadratic & 2048 & 0.34375 & 0.000066 & -0.040192 & 180 & 1.0 \\
    Quadratic & 2048 & 0.375 & 0.000051 & -0.042104 & 180 & 0.30 \\
    Quadratic & 2048 & 0.40625 & 0.000036 & -0.043743 & 180 & 0.50 \\
    Quadratic & 2048 & 0.4375 & 0.000021 & -0.04533 & 180 & 1.0 \\
    Quadratic & 2048 & 0.46875 & 0.000011 & -0.046806 & 180 & 0.20 \\
    Quadratic & 2048 & 0.5 & 0.00001 & -0.049008 & 180 & 1.0 \\
    Quadratic & 2048 & 0.53125 & 0.000025 & -0.046541 & 180 & 0.90 \\
    Quadratic & 2048 & 0.5625 & 0.00004 & -0.04483 & 180 & 1.0 \\
    Quadratic & 2048 & 0.59375 & 0.000055 & -0.042994 & 180 & 0.10 \\
    Quadratic & 2048 & 0.625 & 0.00007 & -0.041105 & 180 & 0.30 \\
    Quadratic & 2048 & 0.65625 & 0.000085 & -0.038943 & 180 & 1.0 \\
    Quadratic & 2048 & 0.6875 & 0.000101 & -0.037186 & 180 & 1.0 \\
    Quadratic & 2048 & 0.71875 & 0.000116 & -0.035207 & 180 & 0.70 \\
    Quadratic & 2048 & 0.75 & 0.000131 & -0.033897 & 180 & 1.0 \\
    Quadratic & 2048 & 0.78125 & 0.000146 & -0.027902 & 180 & 0.80 \\
    Quadratic & 2048 & 0.8125 & 0.000161 & -0.022132 & 180 & 0.60 \\
    Quadratic & 2048 & 0.84375 & 0.000176 & -0.016767 & 180 & 0.10 \\
    Quadratic & 2048 & 0.875 & 0.000191 & -0.011246 & 180 & 1.0 \\
    Quadratic & 2048 & 0.90625 & 0.000207 & -0.005829 & 180 & 0.80 \\
    Quadratic & 2048 & 0.9375 & 0.000222 & -0.000634 & 180 & 0.10 \\
    Quadratic & 2048 & 0.96875 & 0.000237 & 0.004745 & 180 & 0.10 \\
    Quadratic & 2048 & 1 & --- & 0.009438 & --- & --- \\
    \bottomrule
  \end{tabular}
\end{table}

\begin{table}[p]
  \centering
  \caption{\label{tab:supp-benchmark-cubic-32} Complete formation-energy results for the cubic CE Hamiltonian with $N=32$. Column definitions are given in the text.}
  \setlength{\tabcolsep}{16pt}
  \begin{tabular}{cccc ccc}
    \toprule
    Degree & $N$ & $r$ & Weight ($w_r$) & Best Found Objective & Time [s] & Rate of Best \\
    \midrule
    Cubic & 32 & 0 & --- & 0.016659 & --- & --- \\
    Cubic & 32 & 0.03125 & 0.020617 & 0.012218 & 1 & 1.0 \\
    Cubic & 32 & 0.0625 & 0.019017 & 0.00487 & 1 & 1.0 \\
    Cubic & 32 & 0.09375 & 0.017492 & -0.001803 & 1 & 1.0 \\
    Cubic & 32 & 0.125 & 0.015999 & -0.011384 & 1 & 1.0 \\
    Cubic & 32 & 0.15625 & 0.014194 & -0.018705 & 1 & 1.0 \\
    Cubic & 32 & 0.1875 & 0.013081 & -0.028933 & 1 & 1.0 \\
    Cubic & 32 & 0.21875 & 0.024444 & -0.038487 & 1 & 1.0 \\
    Cubic & 32 & 0.25 & 0.010111 & -0.050948 & 1 & 1.0 \\
    Cubic & 32 & 0.28125 & 0.009454 & -0.048397 & 1 & 1.0 \\
    Cubic & 32 & 0.3125 & 0.007648 & -0.049072 & 1 & 1.0 \\
    Cubic & 32 & 0.34375 & 0.006835 & -0.047658 & 1 & 1.0 \\
    Cubic & 32 & 0.375 & 0.004761 & -0.04947 & 1 & 1.0 \\
    Cubic & 32 & 0.40625 & 0.004074 & -0.048703 & 1 & 1.0 \\
    Cubic & 32 & 0.4375 & 0.002795 & -0.051162 & 1 & 1.0 \\
    Cubic & 32 & 0.46875 & 0.003738 & -0.051532 & 1 & 1.0 \\
    Cubic & 32 & 0.5 & 0.000117 & -0.055128 & 1 & 1.0 \\
    Cubic & 32 & 0.53125 & 0.010781 & -0.049538 & 1 & 1.0 \\
    Cubic & 32 & 0.5625 & 0.006308 & -0.046079 & 1 & 1.0 \\
    Cubic & 32 & 0.59375 & 0.003648 & -0.0443 & 1 & 1.0 \\
    Cubic & 32 & 0.625 & 0.002663 & -0.047705 & 1 & 1.0 \\
    Cubic & 32 & 0.65625 & 0.008564 & -0.040148 & 1 & 1.0 \\
    Cubic & 32 & 0.6875 & 0.009539 & -0.034721 & 1 & 1.0 \\
    Cubic & 32 & 0.71875 & 0.0061 & -0.030485 & 1 & 1.0 \\
    Cubic & 32 & 0.75 & 0.005951 & -0.029606 & 1 & 1.0 \\
    Cubic & 32 & 0.78125 & 0.007042 & -0.024317 & 1 & 1.0 \\
    Cubic & 32 & 0.8125 & 0.00833 & -0.020151 & 1 & 1.0 \\
    Cubic & 32 & 0.84375 & 0.008083 & -0.015694 & 1 & 1.0 \\
    Cubic & 32 & 0.875 & 0.009059 & -0.012316 & 1 & 1.0 \\
    Cubic & 32 & 0.90625 & 0.009441 & -0.008263 & 1 & 1.0 \\
    Cubic & 32 & 0.9375 & 0.0104 & -0.005576 & 1 & 1.0 \\
    Cubic & 32 & 0.96875 & 0.010863 & -0.001979 & 1 & 1.0 \\
    Cubic & 32 & 1 & --- & 0.000297 & --- & --- \\
    \bottomrule
  \end{tabular}
\end{table}

\begin{table}[p]
  \centering
  \caption{\label{tab:supp-benchmark-cubic-256} Complete formation-energy results for the cubic CE Hamiltonian with $N=256$. Column definitions are given in the text.}
  \setlength{\tabcolsep}{16pt}
  \begin{tabular}{cccc ccc}
    \toprule
    Degree & $N$ & $r$ & Weight ($w_r$) & Best Found Objective & Time [s] & Rate of Best \\
    \midrule
    Cubic & 256 & 0 & --- & 0.016659 & --- & --- \\
    Cubic & 256 & 0.03125 & 0.002577 & 0.010206 & 10 & 1.0 \\
    Cubic & 256 & 0.0625 & 0.002384 & 0.002475 & 10 & 1.0 \\
    Cubic & 256 & 0.09375 & 0.002207 & -0.005255 & 10 & 1.0 \\
    Cubic & 256 & 0.125 & 0.002034 & -0.014264 & 10 & 1.0 \\
    Cubic & 256 & 0.15625 & 0.001821 & -0.022157 & 10 & 1.0 \\
    Cubic & 256 & 0.1875 & 0.001631 & -0.031328 & 10 & 1.0 \\
    Cubic & 256 & 0.21875 & 0.001487 & -0.040499 & 10 & 1.0 \\
    Cubic & 256 & 0.25 & 0.001334 & -0.050948 & 10 & 1.0 \\
    Cubic & 256 & 0.28125 & 0.001165 & -0.050295 & 10 & 1.0 \\
    Cubic & 256 & 0.3125 & 0.001025 & -0.050371 & 10 & 1.0 \\
    Cubic & 256 & 0.34375 & 0.000855 & -0.050447 & 10 & 1.0 \\
    Cubic & 256 & 0.375 & 0.000682 & -0.051254 & 10 & 1.0 \\
    Cubic & 256 & 0.40625 & 0.000532 & -0.051492 & 10 & 1.0 \\
    Cubic & 256 & 0.4375 & 0.000411 & -0.052461 & 10 & 1.0 \\
    Cubic & 256 & 0.46875 & 0.000309 & -0.053429 & 10 & 1.0 \\
    Cubic & 256 & 0.5 & 0.000135 & -0.055128 & 10 & 1.0 \\
    Cubic & 256 & 0.53125 & 0.000384 & -0.050967 & 10 & 1.0 \\
    Cubic & 256 & 0.5625 & 0.000115 & -0.047821 & 10 & 1.0 \\
    Cubic & 256 & 0.59375 & 0.000186 & -0.048393 & 10 & 1.0 \\
    Cubic & 256 & 0.625 & 0.000374 & -0.04959 & 10 & 1.0 \\
    Cubic & 256 & 0.65625 & 0.000927 & -0.044436 & 10 & 1.0 \\
    Cubic & 256 & 0.6875 & 0.000569 & -0.04003 & 10 & 1.0 \\
    Cubic & 256 & 0.71875 & 0.000641 & -0.035998 & 10 & 1.0 \\
    Cubic & 256 & 0.75 & 0.000768 & -0.03267 & 10 & 1.0 \\
    Cubic & 256 & 0.78125 & 0.000887 & -0.02739 & 10 & 1.0 \\
    Cubic & 256 & 0.8125 & 0.000959 & -0.022695 & 10 & 1.0 \\
    Cubic & 256 & 0.84375 & 0.001055 & -0.018404 & 10 & 1.0 \\
    Cubic & 256 & 0.875 & 0.001136 & -0.014655 & 10 & 1.0 \\
    Cubic & 256 & 0.90625 & 0.001218 & -0.010141 & 10 & 1.0 \\
    Cubic & 256 & 0.9375 & 0.001286 & -0.006211 & 10 & 1.0 \\
    Cubic & 256 & 0.96875 & 0.001355 & -0.002687 & 10 & 1.0 \\
    Cubic & 256 & 1 & --- & 0.000297 & --- & --- \\
    \bottomrule
  \end{tabular}
\end{table}

\begin{table}[p]
  \centering
  \caption{\label{tab:supp-benchmark-cubic-2048} Complete formation-energy results for the cubic CE Hamiltonian with $N=2048$. Column definitions are given in the text.}
  \setlength{\tabcolsep}{16pt}
  \begin{tabular}{cccc ccc}
    \toprule
    Degree & $N$ & $r$ & Weight ($w_r$) & Best Found Objective & Time [s] & Rate of Best \\
    \midrule
    Cubic & 2048 & 0 & --- & 0.016659 & --- & --- \\
    Cubic & 2048 & 0.03125 & 0.000322 & 0.009293 & 180 & 0.10 \\
    Cubic & 2048 & 0.0625 & 0.000298 & 0.001157 & 180 & 0.20 \\
    Cubic & 2048 & 0.09375 & 0.000275 & -0.007254 & 180 & 0.20 \\
    Cubic & 2048 & 0.125 & 0.000252 & -0.015705 & 180 & 1.0 \\
    Cubic & 2048 & 0.15625 & 0.000229 & -0.024156 & 180 & 1.0 \\
    Cubic & 2048 & 0.1875 & 0.000208 & -0.032647 & 180 & 1.0 \\
    Cubic & 2048 & 0.21875 & 0.000186 & -0.041478 & 180 & 1.0 \\
    Cubic & 2048 & 0.25 & 0.000166 & -0.050948 & 180 & 1.0 \\
    Cubic & 2048 & 0.28125 & 0.000146 & -0.050863 & 180 & 1.0 \\
    Cubic & 2048 & 0.3125 & 0.000125 & -0.051142 & 180 & 1.0 \\
    Cubic & 2048 & 0.34375 & 0.000106 & -0.051624 & 180 & 1.0 \\
    Cubic & 2048 & 0.375 & 0.000087 & -0.052146 & 180 & 1.0 \\
    Cubic & 2048 & 0.40625 & 0.000069 & -0.052669 & 180 & 1.0 \\
    Cubic & 2048 & 0.4375 & 0.000051 & -0.053232 & 180 & 1.0 \\
    Cubic & 2048 & 0.46875 & 0.000034 & -0.053997 & 180 & 1.0 \\
    Cubic & 2048 & 0.5 & 0.000018 & -0.055128 & 180 & 1.0 \\
    Cubic & 2048 & 0.53125 & 0.000003 & -0.052478 & 180 & 1.0 \\
    Cubic & 2048 & 0.5625 & 0.000014 & -0.052359 & 180 & 1.0 \\
    Cubic & 2048 & 0.59375 & 0.000031 & -0.049709 & 180 & 0.90 \\
    Cubic & 2048 & 0.625 & 0.000046 & -0.04959 & 180 & 1.0 \\
    Cubic & 2048 & 0.65625 & 0.00006 & -0.046052 & 180 & 1.0 \\
    Cubic & 2048 & 0.6875 & 0.000072 & -0.043209 & 180 & 1.0 \\
    Cubic & 2048 & 0.71875 & 0.000085 & -0.038464 & 180 & 1.0 \\
    Cubic & 2048 & 0.75 & 0.000097 & -0.034202 & 180 & 1.0 \\
    Cubic & 2048 & 0.78125 & 0.000108 & -0.029456 & 180 & 1.0 \\
    Cubic & 2048 & 0.8125 & 0.000119 & -0.025194 & 180 & 1.0 \\
    Cubic & 2048 & 0.84375 & 0.000131 & -0.020448 & 180 & 1.0 \\
    Cubic & 2048 & 0.875 & 0.000141 & -0.016187 & 180 & 1.0 \\
    Cubic & 2048 & 0.90625 & 0.000151 & -0.011627 & 180 & 0.50 \\
    Cubic & 2048 & 0.9375 & 0.000161 & -0.007378 & 180 & 0.10 \\
    Cubic & 2048 & 0.96875 & 0.00017 & -0.003255 & 180 & 0.10 \\
    Cubic & 2048 & 1 & --- & 0.000297 & --- & --- \\
    \bottomrule
  \end{tabular}
\end{table}

\begin{table}[p]
  \centering
  \caption{\label{tab:supp-benchmark-quartic-32} Complete formation-energy results for the quartic CE Hamiltonian with $N=32$. Column definitions are given in the text.}
  \setlength{\tabcolsep}{16pt}
  \begin{tabular}{cccc ccc}
    \toprule
    Degree & $N$ & $r$ & Weight ($w_r$) & Best Found Objective & Time [s] & Rate of Best \\
    \midrule
    Quartic & 32 & 0 & --- & 0.012879 & --- & --- \\
    Quartic & 32 & 0.03125 & 0.020473 & 0.008246 & 1 & 1.0 \\
    Quartic & 32 & 0.0625 & 0.018034 & 0.000907 & 1 & 1.0 \\
    Quartic & 32 & 0.09375 & 0.015783 & -0.005298 & 1 & 1.0 \\
    Quartic & 32 & 0.125 & 0.013779 & -0.013404 & 1 & 1.0 \\
    Quartic & 32 & 0.15625 & 0.011423 & -0.0211 & 1 & 1.0 \\
    Quartic & 32 & 0.1875 & 0.010534 & -0.030516 & 1 & 1.0 \\
    Quartic & 32 & 0.21875 & 0.009966 & -0.038099 & 1 & 1.0 \\
    Quartic & 32 & 0.25 & 0.007459 & -0.047251 & 1 & 1.0 \\
    Quartic & 32 & 0.28125 & 0.007386 & -0.044618 & 1 & 1.0 \\
    Quartic & 32 & 0.3125 & 0.005521 & -0.045861 & 1 & 1.0 \\
    Quartic & 32 & 0.34375 & 0.005205 & -0.045368 & 1 & 1.0 \\
    Quartic & 32 & 0.375 & 0.003294 & -0.047715 & 1 & 1.0 \\
    Quartic & 32 & 0.40625 & 0.003133 & -0.0483 & 1 & 1.0 \\
    Quartic & 32 & 0.4375 & 0.002353 & -0.050492 & 1 & 1.0 \\
    Quartic & 32 & 0.46875 & 0.006863 & -0.052531 & 1 & 1.0 \\
    Quartic & 32 & 0.5 & 0.00035 & -0.056137 & 1 & 1.0 \\
    Quartic & 32 & 0.53125 & 0.009604 & -0.049847 & 1 & 1.0 \\
    Quartic & 32 & 0.5625 & 0.004431 & -0.04749 & 1 & 1.0 \\
    Quartic & 32 & 0.59375 & 0.004451 & -0.045018 & 1 & 1.0 \\
    Quartic & 32 & 0.625 & 0.001124 & -0.04516 & 1 & 1.0 \\
    Quartic & 32 & 0.65625 & 0.010314 & -0.039959 & 1 & 1.0 \\
    Quartic & 32 & 0.6875 & 0.003249 & -0.0387 & 1 & 1.0 \\
    Quartic & 32 & 0.71875 & 0.00513 & -0.034588 & 1 & 1.0 \\
    Quartic & 32 & 0.75 & 0.005756 & -0.0331 & 1 & 1.0 \\
    Quartic & 32 & 0.78125 & 0.014613 & -0.025689 & 1 & 1.0 \\
    Quartic & 32 & 0.8125 & 0.009636 & -0.021412 & 1 & 1.0 \\
    Quartic & 32 & 0.84375 & 0.011018 & -0.01568 & 1 & 1.0 \\
    Quartic & 32 & 0.875 & 0.013303 & -0.012061 & 1 & 1.0 \\
    Quartic & 32 & 0.90625 & 0.01542 & -0.006931 & 1 & 1.0 \\
    Quartic & 32 & 0.9375 & 0.018282 & -0.003368 & 1 & 1.0 \\
    Quartic & 32 & 0.96875 & 0.021142 & 0.003068 & 1 & 1.0 \\
    Quartic & 32 & 1 & --- & 0.007937 & --- & --- \\
    \bottomrule
  \end{tabular}
\end{table}

\begin{table}[p]
  \centering
  \caption{\label{tab:supp-benchmark-quartic-256} Complete formation-energy results for the quartic CE Hamiltonian with $N=256$. Column definitions are given in the text.}
  \setlength{\tabcolsep}{16pt}
  \begin{tabular}{cccc ccc}
    \toprule
    Degree & $N$ & $r$ & Weight ($w_r$) & Best Found Objective & Time [s] & Rate of Best \\
    \midrule
    Quartic & 256 & 0 & --- & 0.012879 & --- & --- \\
    Quartic & 256 & 0.03125 & 0.002566 & 0.006601 & 10 & 0.80 \\
    Quartic & 256 & 0.0625 & 0.002274 & -0.000351 & 10 & 1.0 \\
    Quartic & 256 & 0.09375 & 0.002033 & -0.007594 & 10 & 1.0 \\
    Quartic & 256 & 0.125 & 0.001822 & -0.015047 & 10 & 0.10 \\
    Quartic & 256 & 0.15625 & 0.001538 & -0.022516 & 10 & 1.0 \\
    Quartic & 256 & 0.1875 & 0.001316 & -0.030516 & 10 & 0.80 \\
    Quartic & 256 & 0.21875 & 0.001196 & -0.038527 & 10 & 0.40 \\
    Quartic & 256 & 0.25 & 0.001071 & -0.047251 & 10 & 1.0 \\
    Quartic & 256 & 0.28125 & 0.000916 & -0.047233 & 10 & 0.30 \\
    Quartic & 256 & 0.3125 & 0.000824 & -0.047586 & 10 & 0.30 \\
    Quartic & 256 & 0.34375 & 0.000671 & -0.048125 & 10 & 1.0 \\
    Quartic & 256 & 0.375 & 0.000533 & -0.049562 & 10 & 1.0 \\
    Quartic & 256 & 0.40625 & 0.000436 & -0.050344 & 10 & 1.0 \\
    Quartic & 256 & 0.4375 & 0.000361 & -0.051784 & 10 & 1.0 \\
    Quartic & 256 & 0.46875 & 0.000316 & -0.053555 & 10 & 1.0 \\
    Quartic & 256 & 0.5 & 0.000185 & -0.056137 & 10 & 1.0 \\
    Quartic & 256 & 0.53125 & 0.000261 & -0.052603 & 10 & 1.0 \\
    Quartic & 256 & 0.5625 & 0.000051 & -0.050649 & 10 & 1.0 \\
    Quartic & 256 & 0.59375 & 0.000065 & -0.048894 & 10 & 1.0 \\
    Quartic & 256 & 0.625 & 0.000246 & -0.049829 & 10 & 1.0 \\
    Quartic & 256 & 0.65625 & 0.000697 & -0.045004 & 10 & 1.0 \\
    Quartic & 256 & 0.6875 & 0.000473 & -0.041647 & 10 & 1.0 \\
    Quartic & 256 & 0.71875 & 0.000617 & -0.037643 & 10 & 1.0 \\
    Quartic & 256 & 0.75 & 0.000804 & -0.033455 & 10 & 1.0 \\
    Quartic & 256 & 0.78125 & 0.001019 & -0.027911 & 10 & 1.0 \\
    Quartic & 256 & 0.8125 & 0.001199 & -0.022941 & 10 & 0.10 \\
    Quartic & 256 & 0.84375 & 0.001433 & -0.018079 & 10 & 1.0 \\
    Quartic & 256 & 0.875 & 0.001703 & -0.012581 & 10 & 1.0 \\
    Quartic & 256 & 0.90625 & 0.001998 & -0.007323 & 10 & 0.10 \\
    Quartic & 256 & 0.9375 & 0.002309 & -0.003368 & 10 & 0.20 \\
    Quartic & 256 & 0.96875 & 0.002654 & 0.002676 & 10 & 0.10 \\
    Quartic & 256 & 1 & --- & 0.007937 & --- & --- \\
    \bottomrule
  \end{tabular}
\end{table}

\begin{table}[p]
  \centering
  \caption{\label{tab:supp-benchmark-quartic-2048} Complete formation-energy results for the quartic CE Hamiltonian with $N=2048$. Column definitions are given in the text.}
  \setlength{\tabcolsep}{16pt}
  \begin{tabular}{cccc ccc}
    \toprule
    Degree & $N$ & $r$ & Weight ($w_r$) & Best Found Objective & Time [s] & Rate of Best \\
    \midrule
    Quartic & 2048 & 0 & --- & 0.012879 & --- & --- \\
    Quartic & 2048 & 0.03125 & 0.000322 & 0.006272 & 180 & 0.10 \\
    Quartic & 2048 & 0.0625 & 0.000286 & -0.00089 & 180 & 0.10 \\
    Quartic & 2048 & 0.09375 & 0.000253 & -0.008192 & 180 & 0.10 \\
    Quartic & 2048 & 0.125 & 0.000222 & -0.015864 & 180 & 0.10 \\
    Quartic & 2048 & 0.15625 & 0.000196 & -0.023326 & 180 & 0.10 \\
    Quartic & 2048 & 0.1875 & 0.000173 & -0.031085 & 180 & 0.10 \\
    Quartic & 2048 & 0.21875 & 0.00015 & -0.038963 & 180 & 0.10 \\
    Quartic & 2048 & 0.25 & 0.000132 & -0.047251 & 180 & 0.70 \\
    Quartic & 2048 & 0.28125 & 0.000115 & -0.047542 & 180 & 0.10 \\
    Quartic & 2048 & 0.3125 & 0.000098 & -0.048407 & 180 & 0.60 \\
    Quartic & 2048 & 0.34375 & 0.000084 & -0.049517 & 180 & 1.0 \\
    Quartic & 2048 & 0.375 & 0.00007 & -0.050628 & 180 & 1.0 \\
    Quartic & 2048 & 0.40625 & 0.000058 & -0.051739 & 180 & 1.0 \\
    Quartic & 2048 & 0.4375 & 0.000046 & -0.05285 & 180 & 0.40 \\
    Quartic & 2048 & 0.46875 & 0.000068 & -0.054198 & 180 & 0.20 \\
    Quartic & 2048 & 0.5 & 0.000023 & -0.056137 & 180 & 1.0 \\
    Quartic & 2048 & 0.53125 & 0.000011 & -0.053411 & 180 & 0.40 \\
    Quartic & 2048 & 0.5625 & 0.000001 & -0.052983 & 180 & 1.0 \\
    Quartic & 2048 & 0.59375 & 0.000016 & -0.050257 & 180 & 0.30 \\
    Quartic & 2048 & 0.625 & 0.000031 & -0.049829 & 180 & 1.0 \\
    Quartic & 2048 & 0.65625 & 0.000047 & -0.047085 & 180 & 1.0 \\
    Quartic & 2048 & 0.6875 & 0.000062 & -0.04407 & 180 & 1.0 \\
    Quartic & 2048 & 0.71875 & 0.000162 & -0.03862 & 180 & 0.10 \\
    Quartic & 2048 & 0.75 & 0.000102 & -0.033633 & 180 & 0.70 \\
    Quartic & 2048 & 0.78125 & 0.000126 & -0.028357 & 180 & 0.10 \\
    Quartic & 2048 & 0.8125 & 0.000151 & -0.023018 & 180 & 0.10 \\
    Quartic & 2048 & 0.84375 & 0.000181 & -0.017848 & 180 & 0.10 \\
    Quartic & 2048 & 0.875 & 0.000214 & -0.012519 & 180 & 0.10 \\
    Quartic & 2048 & 0.90625 & 0.000249 & -0.006801 & 180 & 0.10 \\
    Quartic & 2048 & 0.9375 & 0.000289 & -0.001962 & 180 & 0.10 \\
    Quartic & 2048 & 0.96875 & 0.000333 & 0.002949 & 180 & 0.10 \\
    Quartic & 2048 & 1 & --- & 0.007937 & --- & --- \\
    \bottomrule
  \end{tabular}
\end{table}

\clearpage
\begin{figure}[p]
  \centering
  \includegraphics[width=0.72\textwidth]{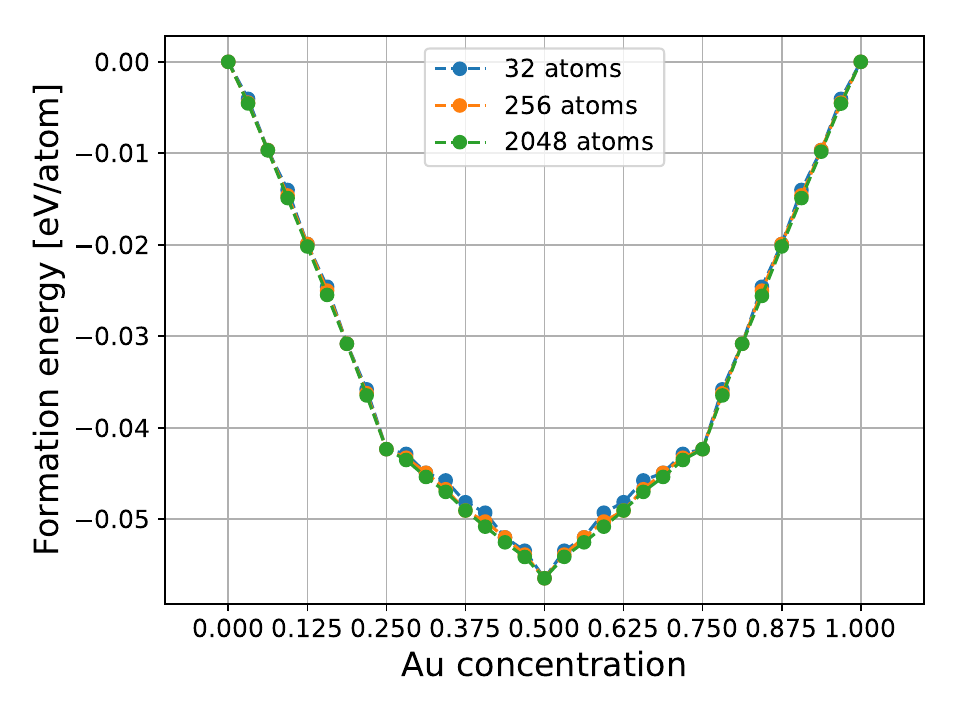}
\caption{
  Formation energy per atom as a function of Au concentration \(r\)
  for configurations obtained with SQBM+ using the quadratic CE
  Hamiltonian. Results are shown for 32-, 256-, and 2048-atom
  supercells. 
  The numerical results underlying these curves are listed in
Tables~S1--S3. A comparison with the corresponding numerical results
from our previous Digital Annealer study is discussed in the main text.
  The principal convex-hull vertices occur at
  \(r=0.25\), \(0.50\), and \(0.75\), corresponding to
  \(L1_2\)-type Cu$_3$Au, \(L1_0\)-type CuAu, and
  \(L1_2\)-type CuAu$_3$ ordering, respectively.
}  \label{fig:formation_energy_quadratic}
\end{figure}
\clearpage

\section{Representative Optimized Structures}
\label{sec:optimized_structures}

\begin{figure}[htbp]
  \centering
  \includegraphics[width=\textwidth]{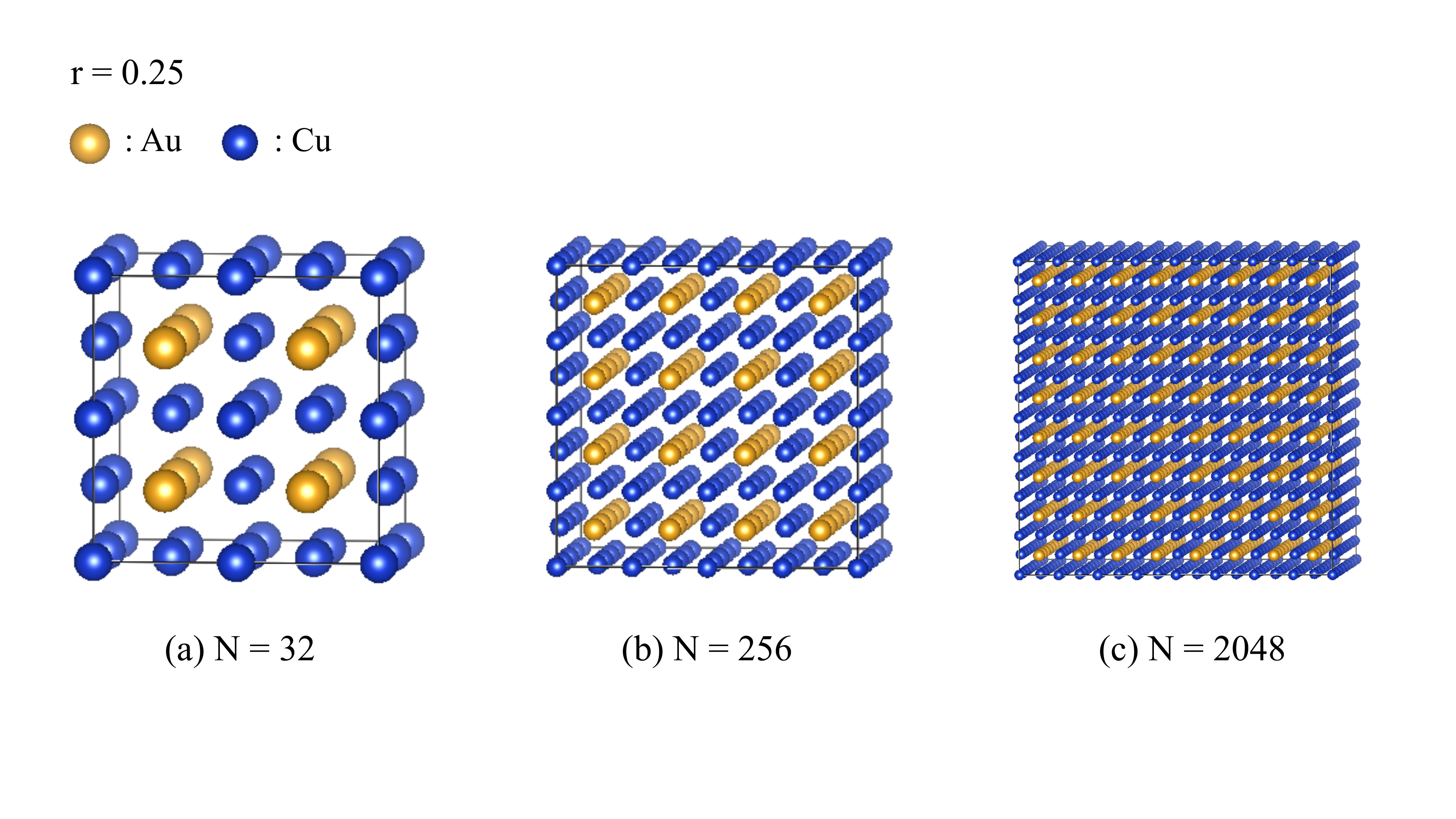}
  \caption{
    Optimized configurations for the quartic CE Hamiltonian at
    $r=0.25$: (a) $N=32$, (b) $N=256$, and (c) $N=2048$.
    The configurations exhibit $L1_2$-type Cu$_3$Au ordering at all
    three system sizes.
  }
  \label{fig:quartic_structure_r025}
\end{figure}

\begin{figure}[p]
  \centering
  \includegraphics[width=\textwidth]{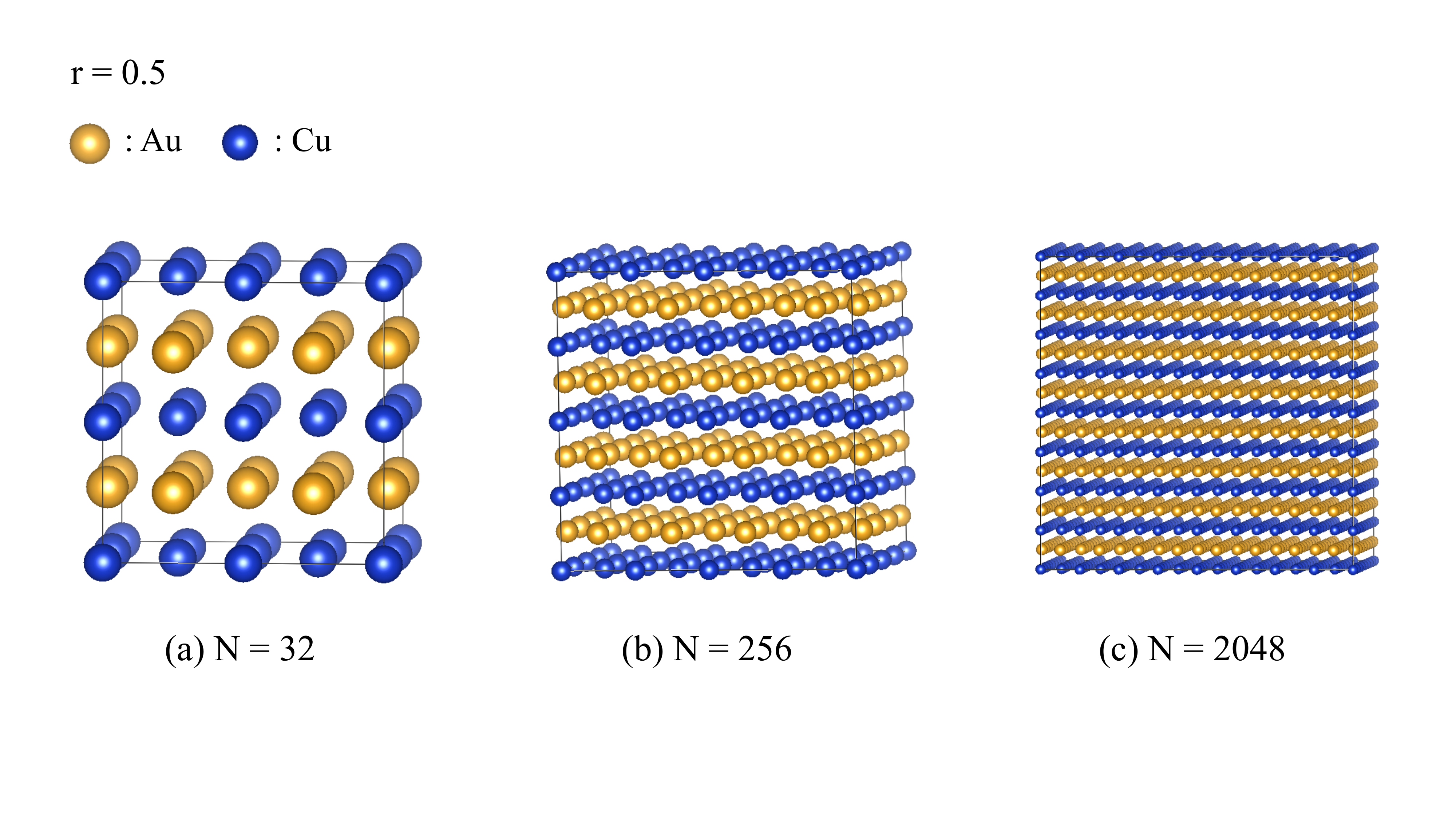}
  \caption{
    Optimized configurations for the quartic CE Hamiltonian at
    $r=0.50$: (a) $N=32$, (b) $N=256$, and (c) $N=2048$.
    The configurations exhibit $L1_0$-type CuAu ordering at all three
    system sizes.
  }
  \label{fig:quartic_structure_r050}
\end{figure}

\begin{figure}[p]
  \centering
  \includegraphics[width=\textwidth]{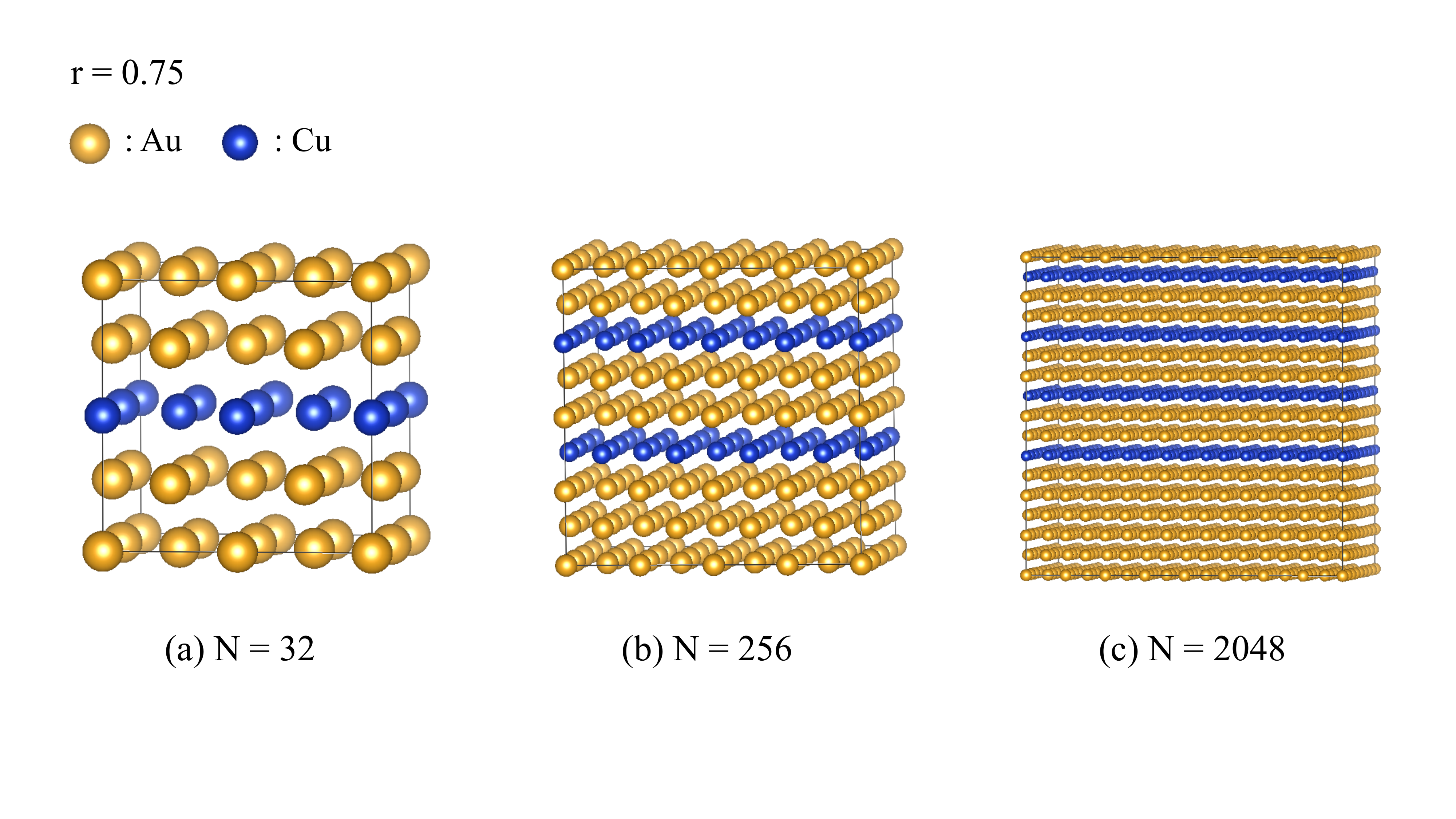}
  \caption{
    Optimized configurations for the quartic CE Hamiltonian at
    $r=0.75$: (a) $N=32$, (b) $N=256$, and (c) $N=2048$.
    The $N=32$ configuration has a $P4/mmm$-type stacking sequence
    comprising one Cu plane and three Au planes. The $N=256$ and
    $N=2048$ configurations are longer-period $P4/nmm$-type layered
    structures with periods of 8 and 16 atomic planes, respectively.
  }
  \label{fig:quartic_structure_r075}
\end{figure}

Figures~\ref{fig:quartic_structure_r025}--\ref{fig:quartic_structure_r075}
show the lowest-objective feasible configurations obtained with SQBM+
for the quartic CE Hamiltonian at three representative compositions.
In all panels, Au and Cu atoms are shown in gold and blue, respectively,
and the outlined boxes indicate the periodic supercells.

\clearpage

\clearpage

\clearpage